



\magnification\magstep1
\baselineskip14pt
\vsize24.0truecm

\input miniltx
\input graphicx

\font\csc=cmcsc10
\font\bigbf=cmbx12
\font\smallrm=cmr8

\def\fermat#1{\setbox0=\vtop{\hsize4.00pc
        \smallrm\raggedright\noindent\baselineskip9pt
        \rightskip=0.5pc plus 1.5pc #1}\leavevmode
        \vadjust{\dimen0=\dp0
        \kern-\ht0\hbox{\kern-4.00pc\box0}\kern-\dimen0}}

\def\section{\bigskip}
\def\subsection{\medskip}
\def\hop{\smallskip\noindent}

\def\rootn{\sqrt{n}}
\def\N{{\rm N}}
\def\E{{\rm E}}
\def\Var{{\rm Var}}

\def\GN{{\rm GN}}
\def\Dir{{\rm Dir}}
\def\tr{{\rm t}}
\def\Tr{{\rm Tr}}

\def\true{{\rm tr}}

\def\diag{{\rm diag}}
\def\arr{\rightarrow}
\def\hatt{\widehat}
\def\tilda{\widetilde}
\def\half{\hbox{$1\over2$}}
\def\eps{\varepsilon}
\def\sumin{\sum_{i=1}^n}
\def\maxin{\max_{i\le n}}
\def\rep{{\rm rep}}
\def\obs{{\rm obs}}
\def\data{{\rm data}}

\def\ppp{{\rm ppp}}
\def\cppp{{\rm cppp}}
\def\c{{\rm c}}
\def\prpp{{\rm prpp}}
\def\midd{\,|\,} 
\def\d{{\rm d}}
\def\Gam{{\rm Gam}}
\def\Bernstein{{Bernshte\u\i n}}
\def\square{{\ \vrule height0.5em width0.5em depth-0.0em}}

\centerline{\bigbf Post-Processing Posterior Predictive P-values}

\medskip
\centerline{\bf Nils Lid Hjort, Fredrik A.~Dahl,  
   and Gunnhildur H\"ognad\'ottir Steinbakk} 
\smallskip
\centerline{\bf University of Oslo}

\smallskip
\centerline{\sl Final version for JASA, October 2005}

{\smallskip\noindent\narrower\baselineskip12pt
{\csc Abstract:}
This article addresses issues of model criticism 
and model comparison in Bayesian contexts, 
and focusses on the use of the 
so-called posterior predictive p-values (ppp values). 
These involve a general discrepancy or conflict measure
and depend on the prior, the model, and the data.
They are used in statistical practice to quantify
the degree of surprise or conflict in data, and for purposes
of comparing different combinations of prior and model.  
The distribution of such ppp values is however 
far from uniform, as we demonstrate for different models, 
making their interpretation and comparison
a difficult matter. We propose a natural calibration
of the ppp values, where the resulting cppp values
are uniform on the unit interval under model conditions. 
The cppp values, which in general rely on a double simulation
scheme for their computation, may then be used to assess 
and compare different priors and models. Our methods also 
make it possible to compare parametric with nonparametric 
model specifications, in that genuine `measures of surprise' 
are put on the same canonical uniform scale. Our techniques 
are illustrated for some applications to real data. 
We also present supplementing theoretical results 
on various properties of the ppp and cppp. 

\smallskip\noindent\sl
{\csc Key words:}
calibration of ppp values, 
dipper data, 
double simulation, 
model criticism, 
posterior predictive p-values,
prior construction,   
prior predictive evaluation, 
surprise
\smallskip}

\section
\centerline{\bf 1. Introduction and summary}

\hop
Bayesian inference involves selecting a prior $\pi(\theta)$ 
for the unknown parameters $\theta$ and a model $f(y,\theta)$
for the data $y$ given $\theta$. In complex situations
some model criticism analysis is often wished for, 
exposing the model to certain posterior checks;
this sometimes leads to modifications in
the model or in the prior. In other situations
there might be a need to consider several candidates 
for both prior and model. This leads to questions on how
to meaningfully criticise, evaluate, compare 
and select among these candidates. 

\subsection
{\sl 1.1. Existing approaches.}
There are by necessity several approaches to handling such 
general problems. 
Classic goodness-of-fit monitoring remains relevant 
(also Bayesians might need to check 
whether data follow a normal distribution). Bayes factors 
(of which there are different related versions) are often used, 
see e.g.~Smith and Spiegelhalter (1980), Kass and Raftery (1995)
and Marden (2000). Related to these again is the so-called 
Bayesian information criterion, the BIC 
(see e.g.~Robert, 2001, Ch.~7, and the lucid discussion
in Clyde and George, 2004). A quite general 
model-evaluation strategy employs the deviance information criterion, 
the DIC, of Spiegelhalter, Best, Carlin and van der Linde (2002)
(see also van der Linde, 2004); 
this method is in widespread use since its computation is 
an easy by-product of the MCMC simulations that are often
used to simulate from the posterior distributions. 
Model adequacy evaluation tools partly geared specifically 
towards use in hierarchical models are proposed and 
discussed in Gelfand and Dey (1994), 
Dey, Gelfand, Swartz and Vlachos (1998), 
O'Hagan (2003), Lu, Hodges and Carlin (2004), 
and Bayarri and Castellanos (2004). 
Versions of the Bayesian model selection problem may
also be cast in decision theoretic terms,
involving utility or loss functions; 
references here include Gelfand and Ghosh (1998),
Guti\'errez-Pe{\~n}a and Walker (2001), 
Claeskens and Hjort (2003), Hjort and Claeskens (2003),
and Kadane and Lazar (2004). 

In addition, various authors have attempted to construct 
`Bayesian p-values', which can be thought of as quantifying
the degree of surprise from data, given the prior and the model,
sometimes also focussing on certain hypotheses. 
The Bayesian p-values come in many forms, and range from 
the prior predictive p-values of Box (1980) to 
the posterior predictive p-values touched on
in Guttman (1967) and Rubin (1984), a tool worked out more 
fully by Gelman, Meng and Stern (1996) and Meng (1994).
Important variations and improvements are introduced 
in Bayarri and Berger (2000), further discussed and analysed 
in Robins, van der Vaart and Ventura (2000),
Bayarri and Berger (2004), and in Bayarri and Castellanos (2004). 

We note that the division line between `model criticism'
(or `model evaluation') and `model comparison' is not sharp; 
if model criticism tools show four out of five models to be 
faulted but accept the fifth, then a comparison is 
already taking place. Nevertheless methods like the BIC
and the DIC are naturally sorted under model comparison,
or model selection, whereas p-value methods, the main
concern of our paper, are meant as model evaluation methods. 

\subsection
{\sl 1.2. The ppp.}
This article focusses on one of the above-mentioned 
mechanisms for carrying out such evaluation and 
comparisons, namely the so-called posterior predictive p-value 
(henceforth, the ppp). It requires specification 
of a suitable discrepancy measure $D=D(y,\theta)$, 
reflecting aspects thought to be important
for the final conclusions of the statistical analysis.
The intention is to assess a posteriori the fit of
the underlying model assumptions, or to quantify 
the degree of surprise by observing what we actually 
observed, in view of prior and model. In the 
formulation of Gelman, Meng and Stern (1996), 
the ppp is defined as 
$$\ppp=\ppp(y^\obs)=\Pr\{D(y^\rep,\theta)
   \ge D(y^\obs,\theta)\midd\data\}. \eqno(1.1)$$
Here $y^\obs$ signifies the observed data, 
or in some cases a suitably relevant subset of the 
full data-set, while $y^\rep$ represents a new (`future') 
data-set of the same type, drawn conjointly with $\theta$ 
from the posterior distribution $\pi(\theta\midd\data)$. 
More concretely, 
$$(\theta,y^\rep)\sim \pi(\theta\midd\data)f(y\midd\theta), \eqno(1.2)$$
so that, in particular, $y^\rep$ comes from the predictive 
distribution $\int f(y\midd\theta)\pi(\theta\midd\data)\,\d\theta$.
Provided we can simulate 
(i) $\theta_j$s from the posterior and
(ii) $y^\rep_j$ data-sets from the model $f(y\midd\theta_j)$,
we may evaluate the ppp as 
$$\ppp(y^\obs)\doteq{1\over A}\sum_{j=1}^A I\{D(y^\rep_j,\theta_j)
   \ge D(y^\obs,\theta_j)\}, \eqno(1.3)$$
across a high number $A$ of simulations. 
It is useful to plot $D(y^\obs,\theta_j)$
vs.~$D(y_j^\rep,\theta_j)$, whereby the ppp number 
is identified as the proportion of points above the diagonal;
see Figure 1.1. 


\centerline{\includegraphics[scale=0.44]{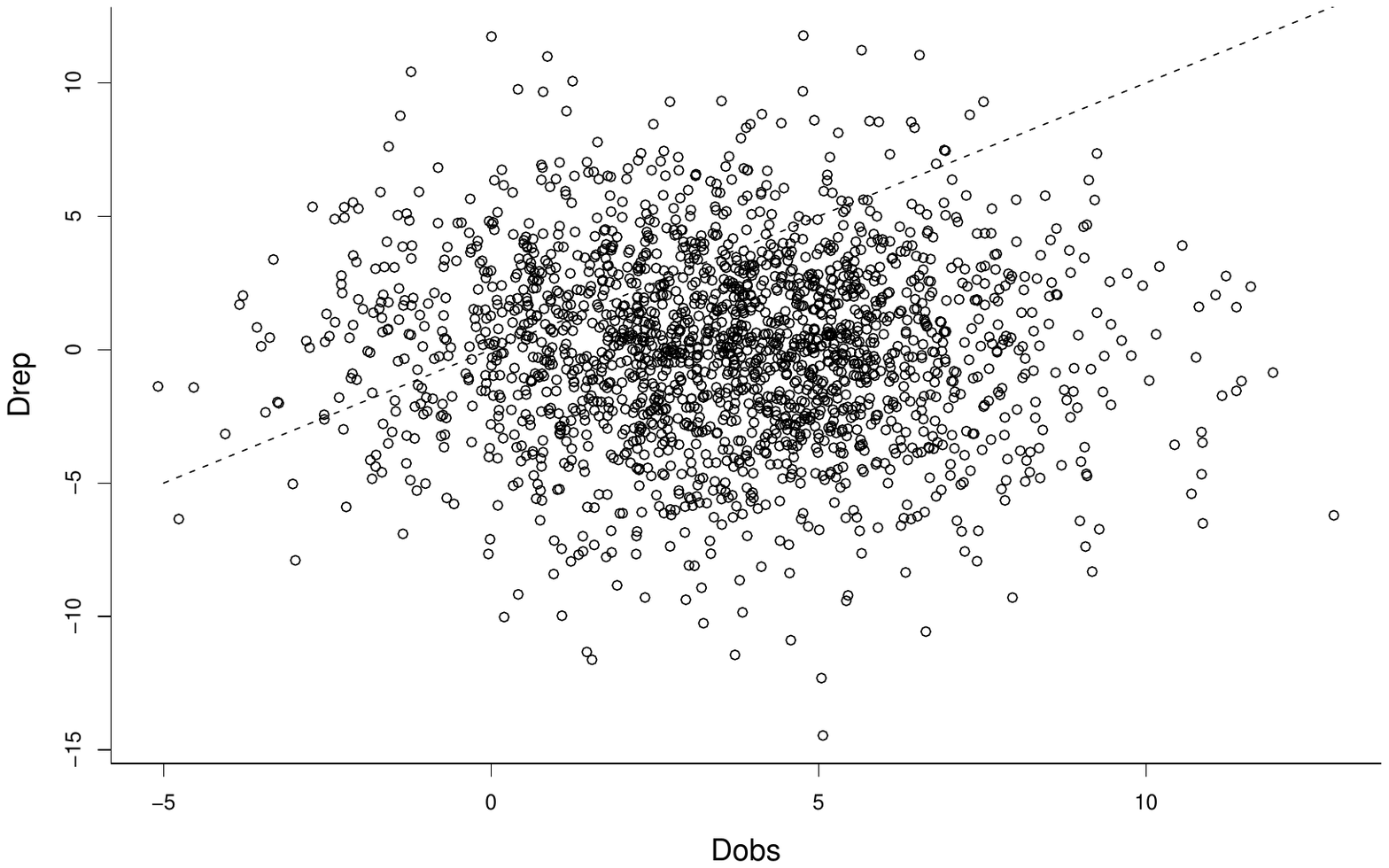}}


\vskip-0.4truecm
{\medskip\sl\narrower\noindent\baselineskip11pt
{\csc Figure 1.1.} 
2000 pairs of $D(y^\rep,\mu,\sigma)$ 
against $D(y^\obs,\mu,\sigma)$ for the speed of light data,
using the normal model with discrepancy measure 
$D=|y_{(61)}-\mu|-|y_{(6)}-\mu|$, 
with a vague uniform prior for $(\mu,\log\sigma)$. 
The $\ppp$ value is .208.
\vskip0pt}

\smallskip
{\sl Example: Newcomb's speed of light data.}
In an impressive 1882 experiment, Simon Newcomb 
attempted to measure the amount of time required 
for light to travel a distance of 7442 meters. 
His 66 observations, recorded as deviations from
24,800 nanoseconds, are used in Gelman, Carlin, Stern and Rubin 
(2004, Section 6.3) to illustrate aspects of the 
$\ppp$ technology. There are two rather extreme values 
at $-44$ and $-2$ with the remaining 64 points 
following a reasonably symmetric distribution 
from 16 to 40 (incidentally, the accepted speed of light 
constant translates to a true value of 33.0 on 
Newcomb's scale). Gelman et al.~first use 
$\min\{y_1,\ldots,y_{66}\}$ as discrepancy  
to show that a $\N(\mu,\sigma^2)$ model does not fit 
the data (there being two clear outliers), 
and then turn to the discrepancy measure 
$D(y,\mu,\sigma)=|y_{(61)}-\mu|-|y_{(6)}-\mu|$, 
where $y_{(i)}$ denotes the $i$th ordered data point. 
They computed $\ppp=.26$ from 200 simulations from
the traditional `objective posterior' for $(\mu,\sigma)$
that corresponds to 
a uniform prior for $(\mu,\log\sigma)$,
and conclude from this that ``any observed asymmetry 
in the middle of the distribution can easily be explained 
by sampling variation''. 
Figure 1.1 displays $(D^\obs,D^\rep)$ pairs 
from 2000 simulations. A precise ppp computation,
with a million rather than 200 or 2000 simulations,
gives $\ppp=.208$.  
\square

\smallskip
The appeal of the ppp apparatus is partly the generous
flexibility afforded the statistician through choices
of the discrepancy function, which can be set up to 
test or inspect different aspects of the model formulation.
Unlike traditional test statistics the discrepancy functions
are also allowed to depend upon the unknown parameters.
Also typically presented as a bonus point is the apparent
lack of strong dependence on the prior; 
the ppp may be computed even with vague 
or objective non-informative priors, as long as 
the posterior is proper. Among the take-home messages
of this article is that the apparent bonus point 
just described is deceiving; even though different
priors may give rise to approximately the same
posteriors, and therefore to approximately the same
ppp numbers, the degree of surprise might be 
quite different. We shall, in particular, argue against
the interpretation implicit in Gelman et al.'s analysis;
we can not know whether $\ppp=.208$ is surprising
or not until one has decided on a proper scale. 
As we shall see, the uniform scale is not at all appropriate. 
 

The ppp value of observed data $y^\obs$ is often thought of
as capable of checking adequacy of a model,
separated from the prior, cf.~the Gelman et al.~quote above.
While this is true for some tailor-made discrepancies, 
it is our view that ppp measures in general 
might better be seen as ways of assessing adequacy 
of the combined (model, prior). 
In the (relatively rare) cases where our prior 
is viewed as `the distribution Nature used when creating the world' 
we might prefer the prior predictive p-value (prpp), 
suggested by Box (1980); this is also a bona fide p-value 
(with null distribution uniform on $[0,1]$). 
Box's prpp can only handle test statistics $D=D(y)$ 
that do not depend on the $\theta$, however. 
Even in cases when the prior is taken quite literally,
therefore, there is a place for ppp values. 

\subsection
{\sl 1.3. The calibrated ppp.}
In a certain sense the ppp calculation uses the data twice; 
first by updating the prior to fit the data better,
and then by estimating how surprising the data are, relative 
to the posterior parameter distribution.
Thus it is not surprising that its distribution, across 
likely values of $y^\obs$, is not uniform; we shall in fact
demonstrate various extreme aspects of non-uniformity
in several situations. This makes the interpretation 
and comparison of ppp values a difficult and risky matter. 
To alleviate this problem our proposal is to post-process 
or calibrate the ppp value, to 
$$\cppp(y^\obs)=\Pr\{\ppp(Y)\le \ppp(y^\obs)\}, \eqno(1.4)$$
where the distribution of $Y$ is that implied by
the prior and model conditions. This cppp, the perfected ppp
value, will now by construction have a uniform null distribution,
i.e.~may be seen as a genuine p-value. The main message
of our article is that the ppp values have limited 
information value in themselves, but that the naturally 
re-scaled cppp versions are genuinely useful and interpretable
across different combinations of priors and models. 

That the ppp numbers sometimes do not convey very useful 
information was not picked up in the initial reactions
when the method was introduced, see e.g.~the discussion
contributions to Gelman, Meng and Stern (1996). There
have been later warnings in the literature, however. 
Dey, Gelfand, Swartz and Vlachos (1998) complained 
that the ppp values were not able to pick out model
inadequacies in a string of Bayesian logistic--normal 
regression setups; Sinharay and Stern (2003) 
similarly observed that the ppp numbers tended to 
cluster too tightly around $\half$ to give clear
signals of model distortions in hierarchical models;
and Bayarri and Berger found $\ppp=.409$ in a situation
where data clash seriously with a model.  
We will in fact demonstrate below that these are 
fairly typical situations; when information content
increases, in relation to the complexity of a model,
the ppp values will tend to cluster around $\half$,
for natural classes of discrepancy functions. 
Our contention is that these problems 
are solved by the calibration mechanism 
alluded to above, as discussed more firmly in Section 4 
and afterwards in our article. When properly calibrated, 
the cppp values are well able to give signals of surprise
or conflict, and may be used to screen away 
unfortunate combinations of prior and data model
in a unified manner. 

Unlike the prior predictive p-value, the ppp value can 
also be applied in cases where the prior is vaguer than
one's `true' subjective belief about the parameters 
in question. The prior may still be informative, but 
does not need to be fine-tuned to reflect all aspects 
of our prior belief. Typically, the ppp value becomes 
less and less dependent on the prior, for a given data model, 
as data information accumulates. The cppp value, however, 
will remain critically dependent on the actual prior used. 
This is at it should be; the calibration transform (1.4) 
is instructed to actively use and assess the implications 
of a given prior.

\subsection
{\sl 1.4. The present article.}
The lay-out of our article is as follows. 
In Section 2 we investigate the structurally simple situation
where data are normal with a normal prior for the mean. 
Here we find a formula for the ppp, and use this to 
study important special cases, corresponding to 
having a sharp prior, a more flat prior, and to 
having a fixed prior with increasing sample size. 
Such calculations and analysis may also be extended to
normal--normal hierarchical models. We also include a brief
large-sample analysis of the ppp for general parametric models. 
Further properties and aspects of the ppp are gleaned 
from studying situations with a mixture of sub-priors, 
in Section 3.
 
Then in Section 4 we propose and develop our perfected ppp
values, the cppp, and discuss their computation and 
interpretation. The cppp may in general be computed 
via a double simulation regime. The details of our 
general theory are then worked out in the context 
of general linear regression models in Section 5, 
illustrated for a real data set, pertaining to 
the modelling of sprint speedskating results, in Section 6.
These techniques could be used routinely in all Bayesian 
analyses of normal-linear models, to check for model 
adequacy and to monitor data for any serious conflicts 
with the prior used. We also illustrate our techniques 
for two models pertaining to survival patterns 
of the European dipper species, in Section 7. 
Brooks, Catchpole and Morgan (2000) have
earlier analysed the same data, using the ppp apparatus,
but via our calibrated ppp we reach somewhat different 
conclusions.

Our cppp values have by construction been transformed 
to a `canonical scale of surprise', namely the uniform
on the unit interval. Observed cppp numbers therefore
enjoy a clear interpretation and can soundly be compared
across several proposed or imagined combinations of 
prior and model. We may even apply the techniques to 
comparison of parametric vs.~nonparametric model 
specifications, as illustrated in Section 8. Such 
comparisons are often quite difficult, conceptually and 
operationally, with other approaches. 

The area of Bayesian model evaluation is evolving rapidly, 
with sometimes conflicting views expressed by different
authors. The reviewers of our manuscript, for example, 
gave somewhat conflicting expert opinions and recommendations
regarding our methods and their ramifications. 
For these reasons we attempt to clarify 
and sum up our main findings and views in Section 9. 
Our article ends in Section 10 with a list of ppp 
and cppp related topics and issues, some worthy 
of further research efforts. 

The success or not of our cppp analysis depends crucially
on the choice of discrepancy measure, the $D(y,\theta)$. 
This choice should be aided by context and aspects of the
actual application, and might in particular be constructed 
to address those model aspects that are seen as crucial
for the principal conclusions of the statistical analysis.
See in this connection constructions of 
Dey, Gelfand, Swartz and Vlachos (1998),
O'Hagan (2003) and Sinharay and Stern (2003). 
Bayesian statisticians need as always to care about
prior, model, and loss function; 
the modern, conscientious members of the species also need to 
exhibit creative acuity for matters pertaining to 
model selection and screening of priors, which 
with the methodology and machinery of this article would mean 
choosing good and problem-relevant discrepancy measures,
followed by appropriate cppp analysis. 

\section
\centerline{\bf 2. The normal--normal model}
   
\hop 
Here we investigate the simple situation where data are normal
with unknown mean and where the prior for this mean parameter 
is also normal. We find an explicit formula for the ppp 
which provides certain general insights into its 
properties and behaviour. Our calculations extend 
to the case with the traditional inverse gamma times normal 
conjugate prior in the general normal model, and may also
be generalised to the case of normal--normal type 
hierarchical models. 

\subsection
{\sl 2.1. A formula for the ppp.}
Assume that the data $y=(y_1,\ldots,y_n)$, 
conditional on $\theta$, are i.i.d.~from $\N(\theta,\sigma^2)$,
with known standard deviation $\sigma$,   
and let the prior be $\theta\sim\N(\theta_0,\sigma_0^2)$. 
We choose to work with 
$$D(y,\theta)=n(\bar y-\theta)^2/\sigma^2
  ={\rm monotone}(|\bar y-\theta|) $$
as discrepancy measure, where $\bar y$ is the mean of $y_i$s;
any monotone increasing function of $|\bar y-\theta|$ gives the
same ppp value, as we see from (1.1). 
This is a pivotal quantity, with distribution 
equal to a $\chi^2_1$ for each given $\theta$. 
For the following result, let $F_{1,1}(v,\kappa)$ be 
the cumulative distribution function of a non-central 
Fisher variable with degrees of freedom $(1,1)$ and
excentricity parameter $\kappa$, i.e.~of a variable
of the form $(X+\kappa^{1/2})^2/Y^2$, where $X$ and $Y$
are independent and standard normals. 

{\smallskip\sl
{\csc Proposition.}
For the situation described, the posterior predictive p-value,
as a function of the observed data, may be expressed as  
$$\ppp(y^\obs)
   =F_{1,1}\Bigl(1+{\sigma^2\over n\sigma_0^2},
   {\sigma^2\over n\sigma_0^2+\sigma^2}
   {(\bar y^\obs-\theta_0)^2\over \sigma_0^2}\Bigr). \eqno(2.1)$$
\smallskip}

{\csc Proof:}
By well-known techniques, the posterior distribution 
for $\theta$ is found to be normal with mean
$(1-\rho_n)\theta_0+\rho_n\bar y^\obs$ and variance
$\rho_n\sigma^2/n$, writing $\rho_n$ for 
$n\sigma_0^2/(n\sigma_0^2+\sigma^2)$. 
Conditional on $\theta$, the $\bar y^\rep$ 
is of course a $\N(\theta,\sigma^2/n)$. 
Write now
$$(\theta\midd\data)
  \sim(1-\rho_n)\theta_0+\rho_n\bar y^\obs+\rho_n^{1/2}(\sigma/\rootn)N_0
  \quad {\rm and} \quad 
  (\bar y^\rep\midd\theta)\sim\theta+(\sigma/\rootn)N, $$
in terms of two independent standard normals $N_0$ and $N$. 
We learn that $D(y^\rep,\theta)=N^2$ for given $\theta$, 
independent of $N_0$, while 
$$D(y^\obs,\theta)={n\over \sigma^2}\Bigl\{(1-\rho_n)(\theta_0-\bar y^\obs)
   +{\sigma\over \rootn}\rho_n^{1/2}N_0\Bigr\}^2
   =\Bigl\{\rho_n^{1/2}N_0
   -(1-\rho_n){\rootn(\bar y^\obs-\theta_0)\over \sigma}\Bigr\}^2. $$
Hence the \ppp{} becomes  
$$\ppp=\Pr\Bigl[N^2\ge \rho_n
   \Bigl\{N_0-{1-\rho_n\over \rho_n^{1/2}}
   {\rootn\over \sigma}(\bar y^\obs-\theta_0)\Bigr\}^2\Bigr], $$
from which the result follows. 
\square 

\subsection
{\sl 2.2. Special cases.}
The formulae above offer insight into the ppp.
Here are some remarks and consequences,
also to be followed up in later sections. 

{\sl Large $n$, or flat prior, or both.}
Suppose $\rootn\sigma_0$ is large, i.e.~either $n$ 
is large, or the prior is flat, or both. 
Then $\rho_n$ goes to 1, and 
$\ppp(y^\obs)\arr\Pr\{N^2\ge N_0^2\}=\half$
for all $y^\obs$. This happens irrespective of the observed 
$\bar y^\obs$, and is illustrated in Figure 2.1, 
for $n=10$, $\sigma=1$, $\theta_0=0$ 
and $\sigma_0=5$, giving $\rho_n=0.996$; 
the ppp is very close to $\half$ over a broad range of $\bar y^\obs$. 

{\sl Moderate $n$ and sharp prior.}
Suppose on the other hand that $\rootn\sigma_0$ 
is small, which means that the prior knowledge 
about $\theta$ is reasonably sharp, compared to 
sample size, i.e.~it is believed rather firmly 
that $\theta$ is close to $\theta_0$.
Then $\rho_n$ is close to zero, and in the limit 
$$\ppp(y^\obs)\arr p^*(\bar y^\obs)
   =\Pr\{N^2\ge n(\bar y^\obs-\theta_0)^2/\sigma^2\}. $$ 
This is the classic p-value for testing the 
hypothesis $\theta=\theta_0$. See Figure 2.1, 
for $n=10$ and $\sigma_0=0.1$, with $\rho_n=0.091$. 
Observe also that for given $\rho_n$, 
there is a maximum attainable value for ppp, namely 
$F_{1,1}(1/\rho_n)$, for $\bar y^\obs=\theta_0$;
this is the `maximally unsurprising' value we 
might have of $\bar y$, under the given prior.  
The ppp distribution is in particular not 
symmetric around $\half$. The figure also displays 
the ppp as a function of $\bar y^\obs$ for 
an intermediate case of $n=10$ and $\sigma_0=1$,
for which $\rho_n=0.909$. We observe that the distribution
of ppp has a clear maximum value just above $\half$
(actually, 0.5151), with most mass from say 0.25 
to this max value. 


\centerline{\includegraphics[scale=0.44]{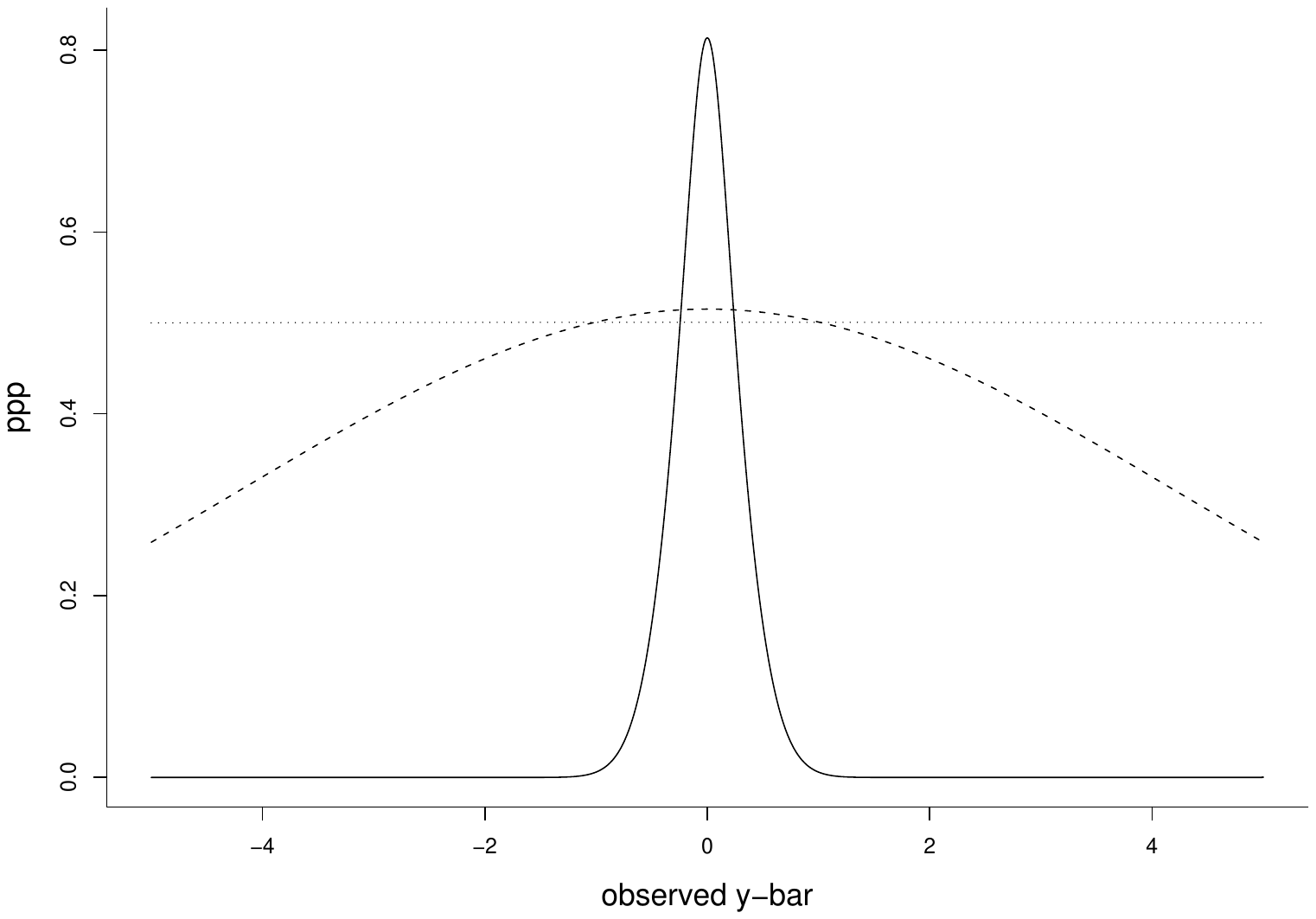}}


\vskip0.1truecm
{\medskip\sl\narrower\noindent\baselineskip11pt
{\csc Figure 2.1.} 
The $\ppp(y^\obs)$ of formula (2.1) is displayed 
as a function of $\bar y^\obs$, 
for $n=10$, $\sigma=1$, $\theta_0=0$, 
for the three cases $\sigma_0$ equal to 
0.1 (solid line), 1.0 (dashed line), 5.0 (dotted line).
\vskip0pt}

\subsection
{\sl 2.3. Full normal analysis.}
Above we took the data standard deviation to be known. 
More realistically, both parameters of the $\N(\mu,\sigma^2)$
distribution are unknown. 
The traditional conjugate prior takes an inverse gamma
for $\sigma^2$ and a normal for $\mu$ given $\sigma$. 
Agree to say that $(\lambda,\mu)=(1/\sigma^2,\mu)$ comes from 
the $\GN(\half a_0,\half b_0,\mu_0,c_0)$ distribution if 
$$\lambda\sim\Gam(\half a_0,\half b_0)
  \quad {\rm and} \quad 
  (\mu\midd\lambda)\sim\N(\mu_0,(c_0\lambda)^{-1}). \eqno(2.2)$$
Its density is accordingly
$$\pi(\lambda,\mu)\propto \lambda^{a_0/2-1}\exp(-\half b_0\lambda)
   \lambda^{1/2}\exp\{-\half\lambda c_0(\mu-\mu_0)^2\}. $$
The likelihood for an observed data set $y_1,\ldots,y_n$ 
from the normal may be written as proportional to 
$\lambda^{n/2}\exp[-\half\lambda\{Q_0^\obs+n(\mu-\bar y^\obs)^2\}]$, 
where $Q_0^\obs=\sumin(y_i-\bar y^\obs)^2$. 
The posterior density becomes
$$\eqalign{
&\propto\lambda^{(a_0+n)/2-1}\lambda^{1/2}
  \exp[-\half\lambda\{b_0+Q_0^\obs+n(\mu-\bar y^\obs)^2+c_0(\mu-\mu_0)^2\}] \cr
&=\lambda^{(a_0+n)/2-1+1/2}
  \exp[-\half\lambda\{b_0+Q_0^\obs
   +(c_0^{-1}+n^{-1})^{-1}(\bar y^\obs-\mu_0)^2 \cr
&\qquad\qquad\qquad\qquad
   +(c_0+n)(\mu-\tilda\mu)^2\}], \cr}$$
proving that 
$$\{(\lambda,\mu)\midd\data\}
  \sim\GN\bigl(\half(a_0+n),\half(b_0+Q_0^\obs
  +(c_0^{-1}+n^{-1})^{-1}(\bar y^\obs-\mu_0)^2),\tilda\mu,c_0+n\bigr), 
   \eqno(2.3)$$
in terms of $\tilda\mu=(c_0\mu_0+n\bar y^\obs)/(c_0+n)$. 
See also the more general analysis of Section 5.

We continue employing the natural discrepancy function 
$D(y,\theta)=n(\bar y-\mu)^2/\sigma^2$, for which, 
when $\theta=(\mu,\sigma)$ is given, $D(y^\rep,\theta)=N^2$ 
for a standard normal $N$, independent of 
$$\eqalign{
D(y^\obs,\theta)
&=n\lambda\{\tilda\mu+(c_0+n)^{-1/2}\lambda^{-1/2}N_0-\bar y^\obs\}^2 \cr
&={n\over c_0+n}\Bigl\{N_0
  +(c_0+n)^{1/2}\lambda^{1/2}{c_0\over c_0+n}
   (-\bar y^\obs+\mu_0)\Bigr\}^2, \cr}$$
where again $N_0$ is standard normal. This yields an exact 
computable expression for the ppp in this case, 
$$\eqalign{
\ppp(y^\obs)
&=\int_0^\infty
   \Pr\Bigl[N^2\ge {n\over c_0+n}\Bigl\{N_0+\lambda^{1/2}
   {c_0\over (c_0+n)^{1/2}}
   (-\bar y^\obs+\mu_0)\Bigr\}^2\midd\lambda\Bigr]\,g_n(\lambda)\,\d\lambda \cr
&=\int_0^\infty F_{1,1}\Bigl({c_0+n\over n},
   {c_0^2\lambda\over c_0+n}(\bar y^\obs-\mu_0)^2\Bigr)\,
   g_n(\lambda)\,\d\lambda, \cr} \eqno(2.4)$$
in terms of the gamma density $g_n(\lambda)$ with 
parameters $(\half a_n,\half b_n)$, where, from the above, 
$a_n=a+n$ and $b_n=b+Q_0^\obs+c_0n(\bar y^\obs-\mu_0)^2/(c_0+n)$. 
The previous case of known $\sigma$ corresponds 
to $a$ and $b$ going to infinity with $a/b=1/\sigma^2$
and $1/(c_0\lambda)=\sigma^2/c_0=\sigma_0^2$, 
i.e.~$c_0=\sigma^2/\sigma_0^2$. 
In this case the above formula reduces to (2.1). 


\smallskip
{\csc Remark 2.1.}
The ppp of (2.1) converges rather rapidly to $\half$
as sample size increases. Some analysis reveals that 
$F_{1,1}(1+a/n,b/n)\doteq \half+\half(a-b)/(\pi n)$, 
with consequences for the (2.1) formula and approximations.
\square

\smallskip
{\csc Remark 2.2.}
The formulae found in Meng (1994, Section 3) relate
to the ppp value for another discrepancy measure,
namely $D(y,\theta_0)=n(\bar y-\theta_0)^2/\sigma^2$,
used as a test statistic for testing $\theta=\theta_0$.
Our ppp measure uses the more general 
$D(y,\theta)$. 
\square

\smallskip
{\csc Remark 2.3.}
Result (2.3) is not new, of course, but the direct 
quick derivation we give takes less journal page 
space than identifying a proper reference 
which then would require careful translation 
of all quantities involved there 
to quantities in our framework. A similar 
comment applies to the more general result
of Section 5.2.
\square

\subsection
{\sl 2.4. ppp behaviour for general parametric models.}
Calculations similar to those above can be carried out
also for more complicated models, with additional efforts.
In work not reported on here we have for example found
explicitly computable formulae for the ppp for the 
situation where data are exponential and the parameter 
has a Gamma prior, as well as for the Poisson--Gamma situation.
In general we would not be able to have explicit formulae
for the $\ppp(y)$, however, and would need to resort
to the simulation scheme of (1.3) to compute the ppp value.

Some general phenomena may be recognised from the
above analysis of the normal--normal and the other cases
mentioned. Suppose that data $y=(y_1,\ldots,y_n)$ 
are i.i.d.~from a density $g(y,\theta)$, conditional 
on $\theta$. Assume also that a discrepancy measure of the type 
$D(y,\theta)=H(\rootn(\hatt\theta(y)-\theta),\theta)$
is used, where $\hatt\theta(y)$ is the maximum likelihood
estimator. We might e.g.~take $H(v,\theta)=v^\tr J(\theta)v$,
with the Fisher information matrix $J(\theta)$; in this case
$D(y,\theta)$ is close to a $\chi^2_p$ for large $n$, 
where $p$ is the dimension of $\theta$. In general,  
$$\ppp(y^\obs)=\Pr\{H(\rootn(\hatt\theta(y^\rep)-\theta),\theta)
   \ge H(\rootn(\hatt\theta(y^\obs)-\theta),\theta)\midd\data\}. $$
Two cases of interest are as follows. 

First consider the case of a `sharp prior', tightly concentrated
around some $\theta_0$. Then this also goes for the posterior, and
$$\ppp(y^\obs)\doteq \Pr\{H(\rootn(\hatt\theta(y^\rep)-\theta_0),\theta_0)
   \ge H(\rootn(\hatt\theta(y^\obs)-\theta_0),\theta_0)\}
   =p^*(y^\obs). $$
This is the classic p-value for testing $\theta=\theta_0$ with
the test $D(y,\theta_0)=H(\rootn(\hatt\theta(y)-\theta_0),\theta_0)$.
This result only requires that $D(y,\theta)$ is continuous 
in $\theta$. 

Then study the large-sample scenario where data 
really follow the $g(y,\theta_\true)$ model, for a suitable
true parameter value, and $n$ grows. 
Under mild regularity conditions,
the distribution of $\rootn(\hatt\theta(y^\rep)-\theta)$
is close to say $V(\theta)$, which is $\N_p(0,J(\theta)^{-1})$,
and this approximation statement holds uniformly 
in a neighbourhood around the $\theta_\true$ value.
In particular, $\rootn(\hatt\theta(y^\rep)-\theta)\arr_d V$,
which is $\N_p(0,J(\theta_\true)^{-1})$.
Secondly, from \Bernstein--von Mises type theorems,
see e.g.~Lehmann (1983, Ch.~6), 
the posterior distribution of $\rootn(\theta-\hatt\theta^\obs)$
is with probability 1 coming close to that of $V_0$, 
another and independent $\N_p(0,J(\theta_\true)^{-1})$ variable. 
All this implies $D(y^\rep,\theta)\arr_d H(V,\theta_\true)$ 
and $D(y^\obs,\theta)\arr_d H(V_0,\theta_\true)$,
with probability 1. As long as $H(v,\theta)$ is continuous,
therefore,  
$$\ppp(y^\obs)\arr\Pr\{H(V,\theta_\true)\ge H(V^0,\theta_\true)\}
   =\half\quad{\rm a.s.} $$
This is the precise description of a phenomenon that 
occasionally has been noted in the literature, 
but perhaps not well understood; 
see e.g.~comments in Sinharay and Stern (2003),
about the ppp values clustering around $\half$. 

\section
\centerline{\bf 3. The ppp when the prior is a mixture} 

\hop 
Here we study the ppp for normal data under a mixture prior
for its mean, and use insights thus revealed to 
make some general comparisons with the so-called 
prior predictive p-values advocated by Box (1980). 

Assume as in the previous section that data $y_1,\ldots,y_n$
conditional on $\theta$ are i.i.d.~from $\N(\theta,\sigma^2)$
and that the same discrepancy measure $D(y,\theta)$ is used, 
but that the prior is a mixture of two different hypotheses
about nature; 
$\theta\sim p_1\N(\theta_{0,1},\sigma_{0,1}^2)
+p_2\N(\theta_{0,2},\sigma_{0,2}^2)$. Then
$$\eqalign{
(\theta,\bar y)
&\sim p_1\pi_1(\theta)f(\bar y\midd\theta)
   +p_2\pi_2(\theta)f(\bar y\midd\theta) \cr
&=p_1\pi_1(\theta\midd\data)f_1(\bar y)
   +p_2\pi_2(\theta\midd\data)f_2(\bar y), \cr}$$
in terms of the posterior densities $\pi_j(\theta\midd\data)$ 
for $\theta$ and of the marginal densities $f_j(\bar y)$
under the two prior hypotheses in question. In fact,
$f_j(\bar y)$ is a normal with mean $\theta_{0,j}$
and variance $\sigma_{0,j}^2+\sigma^2/n$. This leads to 
$$(\theta\midd\data)\sim\tilda p_1(y^\obs)\pi_1(\theta\midd\data)
   +\tilda p_2(y^\obs)\pi_2(\theta\midd\data), $$
where 
$\tilda p_j(y^\obs)=p_jf_j(\bar y^\obs)/
\{p_1f_1(\bar y^\obs)+p_2f_2(\bar y^\obs)\}$ for $j=1,2$. 

This may now be used to find a formula for the $\ppp$.
As in Section 2, $D(y^\rep,\theta)$ may be represented as $N^2$,
where $N$ is a standard normal. For $D(y^\obs,\theta)$, 
it is with probability $\tilda p_j(y^\obs)$ of the type
worked with in Section 2.1, with appropriate parameters
$\rho_{n,j}=n\sigma_{0,j}^2/(n\sigma_{0,j}^2+\sigma^2)$
and $\theta_{0,j}$, for $j=1,2$. Hence
$$\ppp(y^\obs)=\sum_{j=1}^2
   \tilda p_j(y^\obs)F_{1,1}\Bigl({1\over \rho_{n,j}},
   {(1-\rho_{n,j})^2\over \rho_{n,j}}
   {n(\bar y^\obs-\theta_{0,j})^2\over \sigma^2}\Bigr). $$
The formula generalises easily to a mixture across
a wider spectrum of hypotheses about~$\theta$. 

There are several general ppp aspects to be learned 
from applying this formula in different mixture prior settings.
A variety of possible shapes for the $\ppp(\bar y^\obs)$ curves
emerge by different combinations of the parameters. 
Among findings of interest are the following points. 

(i) 
Suppose the two prior hypotheses are both `sharp', 
situated at $\theta_{0,1}$ and $\theta_{0,2}$ with 
small standard deviations $\sigma_{0,1}$ and $\sigma_{0,2}$.
Then we learn that the resulting $\ppp(y^\obs)$ function
is relatively unaffected by the balance parameters 
$p_1,p_2$. The ppp tool essentially works as a classic
frequentist p-value, for testing $\theta=\theta_{0,1}$
if data indicate that it is the first prior component
that is the real one, and for testing $\theta=\theta_{0,2}$
if it is the second prior component that is picked out
by the data. 

For comparison, consider the prior predictive p-value (prpp)
advocated by Box (1980), with respect to the test statistic 
that sorts the $y$ values according to their prior likelihood,
and take $p_2$ very small but positive. 
This $\prpp(y^{\obs})$ would give a value close to zero 
for $\bar y^{\obs}$ close to $\theta_{0,2}$, 
because $\bar y$ values close to $\theta_{0,2}$ 
are highly unlikely under the given prior. 
We may in fact easily prove  
$$\lim_{p_2\arr0}\ppp(\bar y^\obs=\theta_{0,2})=1
  \quad {\rm and} \quad
  \lim_{p_2\arr0}\prpp(\bar y^\obs=\theta_{0,2})=0 $$
in the situation with two sharp hypotheses,
i.e.~$\sigma_{0,j}=0$ for $j=1,2$. 
Thus the ppp may in this respect be seen as a highly 
non-continuous operator, since the two $\ppp(y^\obs)$ 
curves are highly different for $p_2=0$ and $p_2=0.001$. 

(ii) 
Suppose $\sigma_{0,1}$ is moderate while $\sigma_{0,2}$
is small. Then when $\bar y^\obs$ is close to $\theta_{0,2}$
the ppp acts as a classic p-value for testing $\theta=\theta_{0,2}$;
if $\bar y^\obs$ on the other hand is some distance away 
from $\theta_{0,2}$, then the ppp is close to what it 
would be for the first prior component alone. 
Again this behaviour is relatively independent of 
the $p_1,p_2$ parameters. It is interesting and perhaps 
mildly contra-intuitive that even when $p_2$ is very small, 
the ppp indicates non-surprise of $\bar y^\obs$ 
close to $\theta_{0,2}$; thus, in this situation, 
even some events that have very low prior probability 
are deemed completely acceptable. See Figure 3.1.

(iii) 
Let us now hold $p_2$ fixed (but small), while $\theta_{0,2}$ varies.
We have seen that when $\theta_{0,2}$ is large, 
$\ppp(\bar y^\obs=\theta_{0,2})$ will be close to 1.
Also, if $\theta_{0,2}\arr0$, we clearly get 
$\ppp(\bar y^\obs=\theta_{0,2})\arr1$.
But if $\theta_{0,2}$ is at a moderate distance from 
$\theta_{0,1}$, then $\ppp(\bar y^\obs=\theta_{0,2})$ 
may be close to zero. This is a bit of a paradox: 
we have two competing models, represented through 
a prior with two peaks, and observe a value that fits 
the a priori unlikely model well. If the models 
are either very different or very similar, then 
ppp gives a high value, while if they are moderately 
different, we get a low ppp value.


\centerline{\includegraphics[scale=0.44]{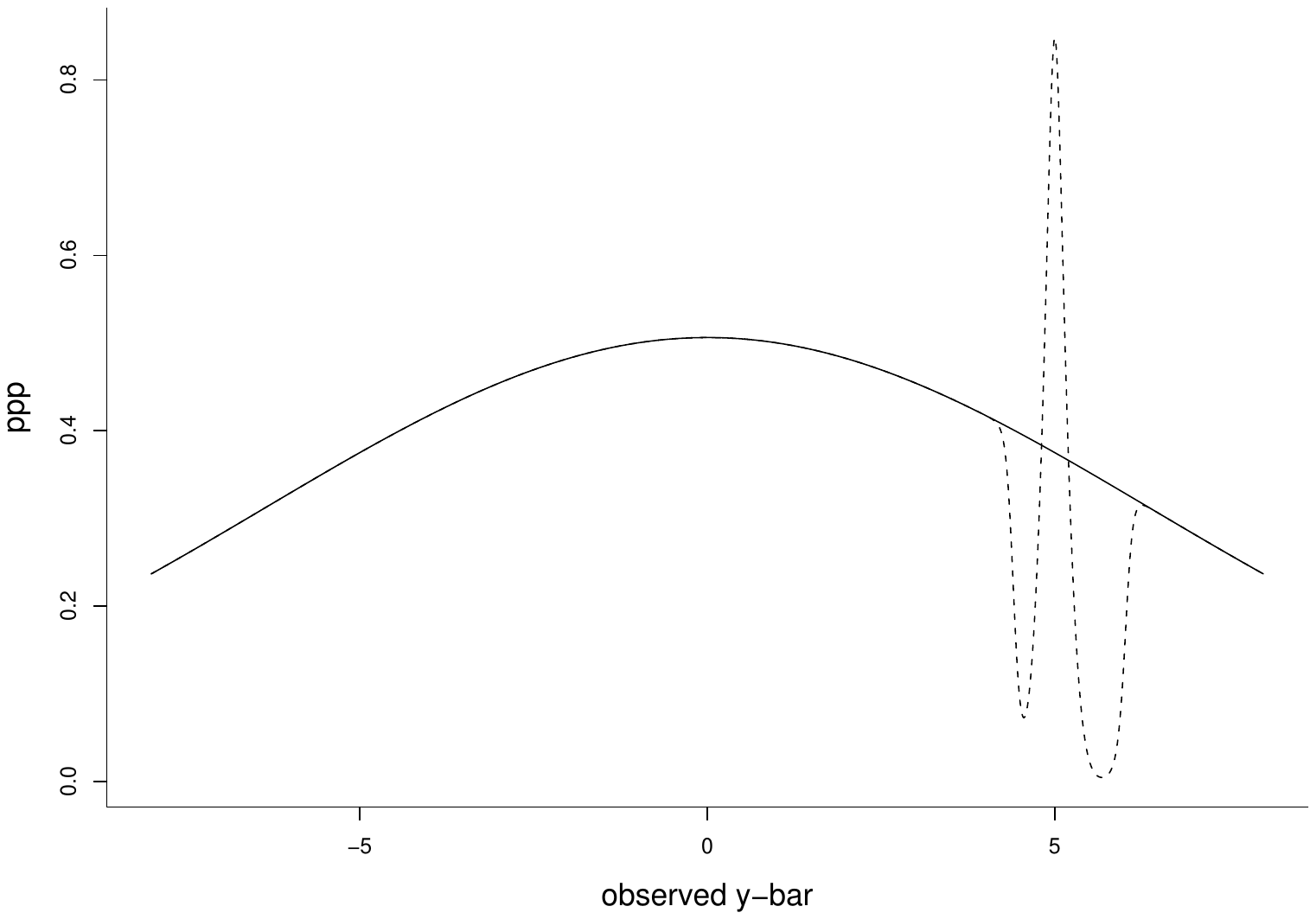}}


\vskip0.1truecm
{\medskip\sl\narrower\noindent\baselineskip11pt
{\csc Figure 3.1.} 
The $\ppp(y^\obs)$ as a function of $\bar y^\obs$, 
for $n=25$ and $\sigma=1$, displayed for two priors;
for the normal $(0,1)$ (solid line) and 
for a sharp bi-mixture with $p_1=0.999$ and $p_2=0.001$,
of two normals $(0,1)$ and $(5,0.05^2)$ (dotted line). 
\smallskip}

These examples illustrate the fact that the ppp value is 
relatively insensitive to the magnitude of the probability 
mass that the prior assigns to parts of the parameter space 
that are distant from each other. This is in particular true 
if the different parameter values give very different
distributions for the observed $y^\obs$.
If this is the case, conditioning on the observation will 
essentially eliminate parameter values that fail to explain 
the data, making their prior likelihood irrelevant.


\section
\centerline{\bf 4. ppp calibration and the double-level simulation} 

\hop 
Suppose one computes $\ppp(y^\obs)=.28$ for one's data set, 
for a given prior and model. This is 
a well-defined probability, as in (1.1), but to
judge the significance of the .28 number, possibly
in comparison with other ppp values for other combinations
of prior and model for the same data set, we are forced
to define the underlying probability scale: how rare,
or how common, are values less than .28? 

There has perhaps been a certain Pavlovian tendency 
in applied statistics work to 
interpret the ppp numbers on the uniform scale, 
like for classic p-values. We have already seen 
that the distribution of ppp is often 
quite non-uniform, however, and more precise information
is provided below. We shall see that an observed $\ppp=.28$
(say) may be extremely surprising in one situation, 
whereas the value $\ppp=.06$ (say) may not be 
very surprising in another situation. 
Like frequentists, who compute p-values as probabilities
of events involving outcomes that did not occur, 
here also the Bayesian (if a priori willing to 
work with the ppp in the first place) is forced 
to consider the values of $y$ that one did not observe. 
In this section we define and study the appropriate
distribution of $\ppp(Y)$ across values of $Y$,
and use this to introduce our calibrated ppp value,
the $\cppp(y^\obs)$. Its actual use is illustrated with
real data in Sections 6 and 7. 

\subsection
{\sl 4.1. The null-null distribution of the ppp.}
Classical Bayes analysis operates of course conditional 
on $y^\obs$. When one in addition wishes to test validity 
of aspects of model and prior, however, one needs to 
take into account also the distribution of $y^\obs$. 
This motivates defining `the null-null distribution' of 
$\ppp(Y)$, corresponding to the distribution of $\ppp(y)$ 
across precisely those $y$ values that occur by the
combined mechanism of the prior and the model. 
The null-null distribution of $U=\ppp(Y)$, 
under perfect prior and perfect model, is
$$G(u)=\Pr\{\ppp(Y)\le u\} \quad {\rm where\ }
   Y\sim\int f(y,\theta)\pi(\theta)\,\d \theta. \eqno(4.1)$$
This is the distribution that should be used to calibrate
the observed ppp value. We propose using 
$$\cppp(y^\obs)=G(\ppp(y^\obs))
   =\Pr\{\ppp(Y)\le\ppp(y^\obs)\}. \eqno(4.2)$$
The distribution of $\cppp(Y)$, across values of $Y$ 
as above, is then by construction a uniform on $(0,1)$
(as long as $G$ is continuous). 
In other words, the cppp is a proper p-value. 

To illustrate, consider the normal--normal situation
of Section 2, and suppose the world is exactly as we have 
imagined, both regarding Nature's distribution of $\theta$ 
and our own modelling of data given $\theta$. Then 
$\bar Y$ is normal $(\theta_0,\sigma_0^2+\sigma^2/n)$, 
expressible as $\theta_0+(\sigma_0^2+\sigma^2/n)^{1/2}M$ 
with a standard normal $M$. 
The consequent null-null distribution of ppp becomes
$$\eqalign{
G(u)
&=\Pr\Bigl\{F_{1,1}\Bigl({1\over \rho_n},{(1-\rho_n)^2\over \rho_n}
   {n(\bar Y-\theta_0)^2\over \sigma^2}\Bigr)\le u\Bigr\} \cr
&=\Pr\Bigl\{{(1-\rho_n)^2\over \rho_n}
   {n(\bar Y-\theta_0)^2\over \sigma^2}\ge q(u)\Bigr\} \cr
&=\Pr\Bigl\{{n\over \sigma^2}\Bigl(\sigma_0^2+{\sigma^2\over n}\Bigr)M^2
   \ge{\rho_n\over (1-\rho_n)^2}q(u)\Bigr\}
 =\Pr\{\chi^2_1\ge q(u)\rho_n/(1-\rho_n)\} \cr}$$
for $0\le u\le F_{1,1}(1/\rho_n)$. 
Here $q(u)$ is the excentre parameter that makes 
$F_{1,1}(1/\rho_n,q)=u$. 


\centerline{\includegraphics[scale=0.44]{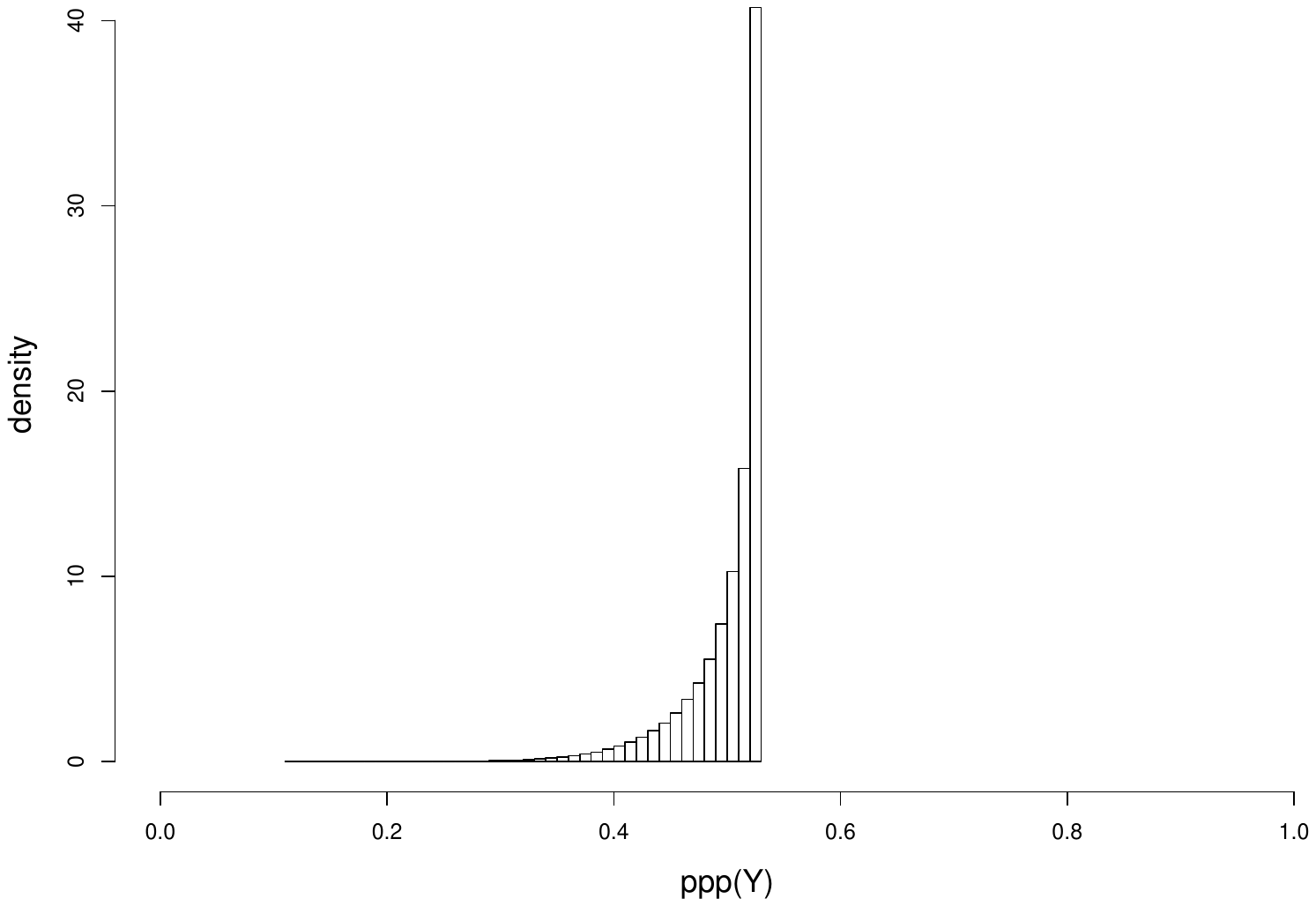}}


\vskip0.1truecm
{\medskip\sl\narrower\noindent\baselineskip11pt
{\csc Figure 4.1.} 
Density of $\ppp(Y)$, displayed as a normalised
histogram with a million simulations, 
for $n=5$, $\sigma=1$, $\sigma_0=1$. 
The distribution has exact mean~$\half$, 
with sharp right cut-off point at $F_{1,1}(1/\rho_n)$.
\smallskip}

The ppp distribution can also easily be displayed 
via simulation of 
$$U=F_{1,1}\Bigl({1\over \rho_n},
   {(1-\rho_n)^2\over \rho_n}(1+n\sigma_0^2/\sigma^2)M^2\Bigr)
   =F_{1,1}\Bigl({1\over \rho_n}, 
   {1-\rho_n\over \rho_n}M^2\Bigr), \eqno(4.3)$$
where $M\sim\N(0,1)$.  
To compute the $\cppp(y^\obs)$ we may simply check the
relative frequency of such simulated $U$s that fall below 
the observed $\ppp(y^\obs)$. For this particular situation
we may even get an explicit expression; 
$$\eqalign{
\cppp(y^\obs)
&=\Pr\Bigl\{F_{1,1}\Bigl({1\over \rho_n},{(1-\rho_n)^2\over \rho_n}
   {n(\bar Y-\theta_0)^2\over \sigma^2}\Bigr)
  \le F_{1,1}\Bigl({1\over \rho_n},{(1-\rho_n)^2\over \rho_n}
   {n(\bar y^\obs-\theta_0)^2\over \sigma^2}\Bigr)\Bigr\} \cr
&=\Pr\{n(\bar Y-\theta_0)^2/\sigma^2\ge n(\bar y^\obs-\theta_0)^2/\sigma^2\}\cr
&=\Pr\Bigl\{\chi^2_1\ge
   {n(\bar y^\obs-\theta_0)^2/\sigma^2
   \over 1+n\sigma_0^2/\sigma^2}\Bigr\}. \cr} \eqno(4.4)$$ 

While the classic p-value statistic is uniform 
on the unit interval under the null hypothesis, 
the present null-null distribution is quite far from 
having such a form. The $U$ variable of (4.3)
is confined to $[0,F_{1,1}(1/\rho_n)]$, with a sharp 
upper threshold, and is highly skewed to the left. 
For $\rootn\sigma_0$ moderate or large 
(sample size is moderate, or prior is moderately flat, or both),
$\rho_n$ is close to 1 and the ppp distribution 
is quite tightly concentrated around $\half$.
Only for $\rootn\sigma_0$ quite small, with $\rho_n$ 
close to zero, does the ppp distribution come close 
to the uniform one on the unit interval (which is 
the limit case as $\rho_n\arr0$). The statistical 
distribution of $\cppp(Y)$ of (4.4), however, 
is by construction exactly a uniform on $[0,1]$,
provided the prior and the model are correct. 
Figure 4.1 shows the partly extreme nature of the ppp distribution,
while Figure 4.2 gives the ppp and cppp curves,
along with a third variant treated in a later subsection. 

\def\conf{{\rm conf}}

\smallskip
{\csc Remark 4.1.}
When $n$ grows, the observed $\bar y^\obs$ will converge 
to the true underlying mean value $\theta_\true$, 
and from (4.4) we see that $\cppp(y^\obs)$ 
tends a.s.~to $\Pr\{|\N(0,1)|\ge |\theta_\true-\theta_0|/\sigma_0\}$. 
Thus 
$$\conf=|\theta_\true-\theta_0|/\sigma_0
   =c_0^{1/2}|\theta_\true-\theta_0|/\sigma $$
emerges as a natural measure of conflict between the real data 
distribution and the $\N(\theta_0,\sigma_0^2)$ prior. 
If $\conf\ge1.96$, then $\cppp(y^\obs)$ will for large $n$
be below the critical value 0.05, etc. 
The second representation of $\conf$ uses $c_0=\sigma^2/\sigma_0^2$, 
which has interpretation as prior sample size. 
See Remark 5.1 for a generalisation and Section 10.1
for a constructive prior-calibration idea.  
\square


\subsection
{\sl 4.2. The double simulation method to calibrate ppp.}
The reasoning above invites the following simulation
method to compute the calibrated ppp value. Simulate
values $(\theta_k,y_k)$ for $k=1,\ldots,B$, for a high
number $B$, where $\theta_k\sim\pi(\theta)$ and 
the full data set $y_k$ is drawn from the model given $\theta_k$.
Then compute 
$$\cppp(y^\obs)\doteq{1\over B}\sum_{k=1}^B
   I\{\ppp(y_k)\le\ppp(y^\obs)\}. \eqno(4.5)$$
It is the perhaps chief claim of our article that 
while the $\ppp(y^\obs)$ of (1.3) may have a difficult 
interpretation and sometimes a low information value, 
the $\cppp(y^\obs)$ of (4.5) has a clear meaning
and can be highly informative. 

While clear in interpretation and natural qua strategy, 
the (4.5) operation might of course be both cumbersome and
computer time costly from an operational point of view,
since it in general must amount to a double simulation,
with $AB$ operations in total, following (1.3). 
One should therefore look for ways of simplifying
the computational burden. Sometimes an explicit formula
may be worked out for the $\ppp(y^\obs)$, with 
considerable benefit for the $\cppp$ computations, 
as we also see in the next section, or one may
be helped by knowing the distribution of $D(y^\rep,\theta)$. 
In other cases one may look for variance reduction tricks 
or for ways of approximating the $\ppp(Y)$ distribution,
the benefit being that one may be allowed a moderate
rather than a large number $B$ of repeated sampling. 
We have actually developed some methods of this kind 
in connection with the cppp analysis of the 
bird survival data in Section 7, but this will 
be reported on elsewhere. 

\def\calM{{\cal M}}

\subsection
{\sl 4.3. Alternative what-if scenarios.}
We judge the above calibration to be the canonical one, 
transforming the ppp numbers under the proposed prior 
and model to the uniform scale. One may however also consider
other what-if scenarios that from different perspectives
lead to other potentially interesting distributions for $Y$,
and hence to other calibrations for $\ppp(Y)$,
in the formulation of (4.1)--(4.2). Suppose in general terms 
that a distribution for such $Y^*$ data-sets is being
considered, where such $Y^*$ are drawn from a mechanism
different from the canonical one given in (4.1); this
other distribution could for example take the form 
$\int f(y,\theta)\pi^*(\theta)\,\d\theta$, for a prior
$\pi^*(\theta)$ different from $\pi(\theta)$. This defines
an alternative ppp distribution $G^*(u)=\Pr\{\ppp(Y^*)\le u\}$
and in its turn a differently calibrated ppp value, 
$$\cppp^*(y^\obs)
   =G^*(\ppp(y^\obs))
   =\Pr\{\ppp(Y^*)\le\ppp(y^\obs)\}. \eqno(4.6)$$
A ppp number is computed under a given model specification,
say $\calM$, and we may write $\ppp(y^\obs,\calM)$ 
to indicate this. Whereas we above calibrated 
$\ppp(y^\obs,\calM)$ via the same model $\calM$, 
operations as described earlier amount to 
calibrating $\ppp(y^\obs,\calM)$ using 
an alternative model $\calM^*$ for data-sets $Y^*$. 
The (4.6) scheme may then be thought of in terms of  
$$\eqalign{
\cppp^*(y^\obs)
&=\c[\ppp(y^\obs,\calM),\calM^*] \cr
&={\rm calibrated\ }\ppp(y^\obs,{\rm model\ }\calM),
   {\rm under\ model\ }\calM^*. \cr} \eqno(4.7)$$

The definition here is fully general and operational, 
and the $\cppp^*$ may be computed for any well-defined 
$Y^*$ distribution, via double simulation if necessary. 
For the normal--normal model we may see more 
clearly the implications of the (4.6) idea; 
the reasoning that led to (4.4) now gives  
$$\cppp^*(y^\obs)=\Pr\{n(\bar Y^*-\theta_0)^2/\sigma^2
   \ge n(\bar y^\obs-\theta_0)^2/\sigma^2\}, \eqno(4.8)$$  
in terms of the distribution of the sample average $\bar Y^*$ 
stemming from data-sets $Y^*$ drawn from the intended 
alternative what-if distribution. This gives formulae
generalising that of (4.4), for different models $\calM^*$
for data-sets $Y^*$. 

\def\prior{{\rm prior}}
\def\posterior{{\rm posterior}}


\centerline{\includegraphics[scale=0.44]{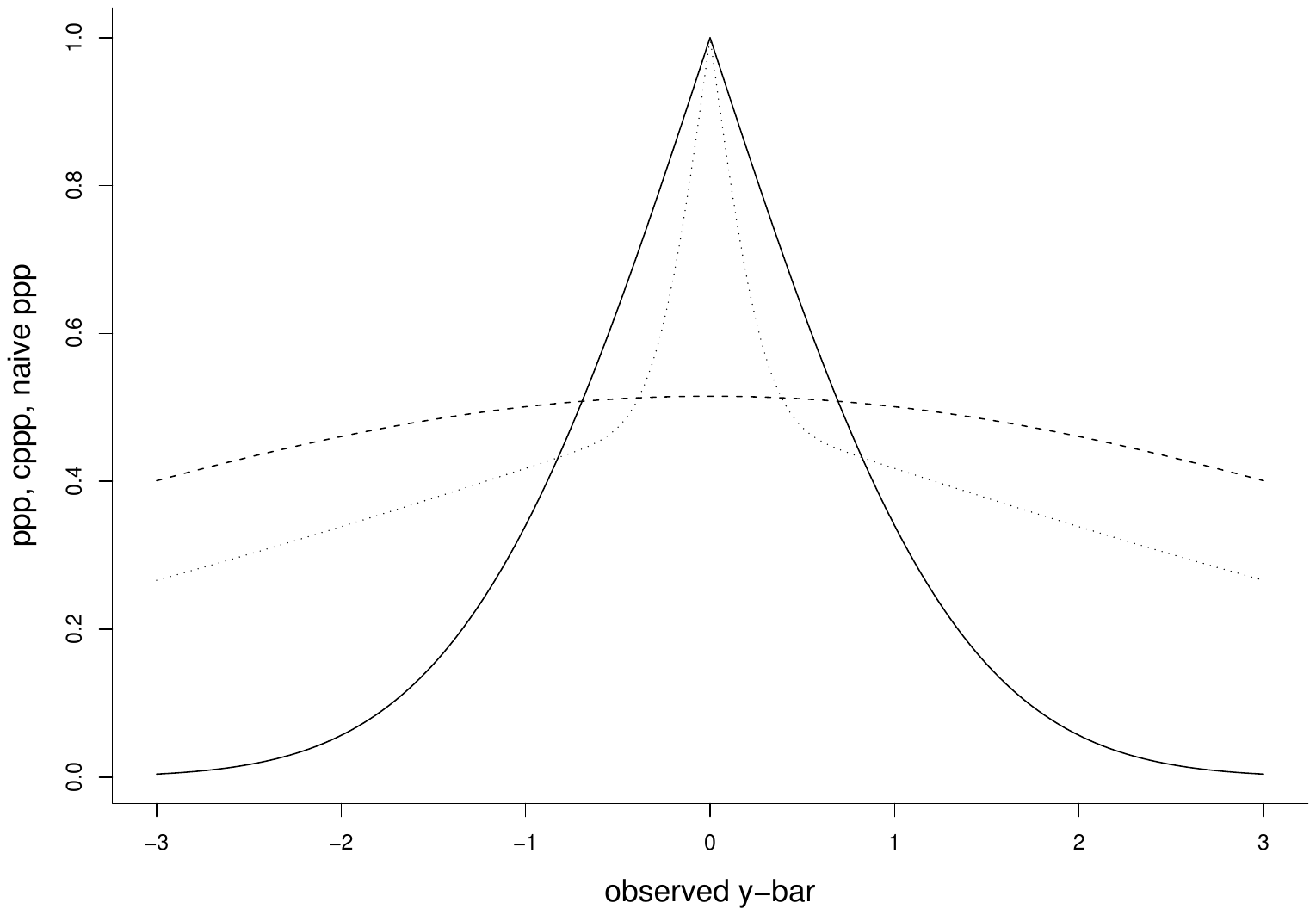}}


\vskip0.1truecm
{\medskip\sl\narrower\noindent\baselineskip11pt
{\csc Figure 4.2.} 
The three functions $\ppp$ (dashed line), 
$\cppp$ (solid line), $\cppp^*$ (dotted line)
are displayed as functions of $\bar y^\obs$,
for $n=10$, $\sigma=1$, with a $\N(0,1)$ prior
for $\theta$. 
\smallskip}

As a particular illustration we may consider the somewhat
naive data-twice-version that instead of sampling 
$y$ from the prior predictive $\int f(y,\theta)\pi(\theta)\,\d \theta$
uses the posterior predictive 
$\int f(y,\theta)\pi(\theta\midd\data)\,\d\theta$.
This has some intuitive attraction, in that it does 
take on board the new knowledge about $\theta$ that 
was not available before data; it would also be 
close to using $f(y,\hatt\theta)$ as the model 
for data, with the maximum likelihood estimator. 
There are instances in the literature of such 
analysis, sometimes carried out tentatively to illustrate
consequences under different scenarios. In Bayesian 
clinical trials literature one sometimes discusses
`sampling density' vs.~`fitting density', for example;
see Gelfand and Wang (2002) for some related discussion. 

For the normal--normal situation, once more, 
we know that $\theta\midd\data$ is distributed as a 
normal with mean $(1-\rho_n)\theta_0+\rho_n\bar y^\obs$
and variance $\rho_n\sigma^2/n$, with $\rho_n$ 
as in Section 2.1. Sampling data $Y_i^*$
from the $\N(\theta,\sigma^2)$ for such $\theta$ 
we find that 
$\bar Y^*-\theta_0$ is normal 
$(\rho_n(\bar y^\obs-\theta_0),(1+\rho_n)\sigma^2/n)$. 
Using the $\cppp^*$ formula (4.8) this yields
$$\eqalign{
\cppp^*(y^\obs)
&=\Pr\bigl[\{\rho_n(\bar y^\obs-\theta_0)
   +(1+\rho_n)^{1/2}(\sigma/\rootn)N\}^2
   \ge (\bar y^\obs-\theta_0)^2\bigr] \cr
&=\Pr\Bigl[{\sigma^2\over n}(1+\rho_n)
   \Bigl\{{\rootn\over \sigma}
   {\rho_n\over (1+\rho_n)^{1/2}}
   (\bar y^\obs-\theta_0)+N\Bigr\}^2
   \ge (\bar y^\obs-\theta_0)^2\Bigr] \cr
&=\Pr\Bigl\{\chi^2_1\Bigl({\rho_n^2\over 1+\rho_n}z_n^\obs\Bigr)
   \ge {z_n^\obs\over 1+\rho_n}\Bigr\}, \cr}$$
in terms of $z_n^\obs=n(\bar y^\obs-\theta_0)^2/\sigma^2$. 
In the terminology and thinking of (4.7) this would 
be the formula for 
$$\eqalign{
\cppp^*(y^\obs)
&=\c[\ppp(y^\obs,\prior),\posterior] \cr
&={\rm calibrated\ }\ppp(y^\obs,{\rm model\ }\prior),
   {\rm under\ model\ }\posterior. \cr}$$
Our previous $\cppp(y^\obs)$ is in this terminology
the same as $\c[\ppp(y^\obs,\prior),\prior]$. 

Figure 4.2 illustrates the $\ppp$, $\cppp$ and $\cppp^*$
curves for $n=10$, $\sigma=1$, with a normal $(0,1)$
prior for $\theta$, as a function of $\bar y^\obs$. 
We see that the $\cppp^*$ is not a satisfactory measure 
of surprise; we would scarcely ever be surprised,
if the numbers are interpreted on the uniform scale.
This is due to the double use of data when we 
draw $Y^*$ from the posterior predictive.  
There could be other choices for $Y^*$ distributions
with more relevance and with a less over-cautious
$\cppp^*$ curve than here. 

\section
\centerline{\bf 5. ppp and cppp analysis for general regression}

\hop 
Here we study the general linear regression model for 
data $(x_i,y_i)$ for which $y_i=x_i^\tr\beta+\eps_i$,
for $i=1,\ldots,n$, where $x_i$ is a $p$-dimensional
covariate vector for individual $i$, and the $\eps_i$s
are independent and zero-mean normal with standard 
deviation $\sigma$. In standard matrix formulation,
$y=X\beta+\eps$, with least squares estimator 
$\hatt\beta=(X^\tr X)^{-1}X^\tr y$ for $\beta$;
it is assumed that the $n\times p$ matrix $X$ is of full rank.
We shall develop theory for a canonical ppp measure 
that makes a good quality evaluation of the underlying
model assumptions. It is based on the discrepancy
measure 
$D(y,\theta)=(\hatt\beta-\beta)^\tr\Omega_n(\hatt\beta-\beta)/\sigma^2$,
where $\Omega_n=X^\tr X=\sumin x_ix_i^\tr$, and where we write 
$\theta=(\beta,\sigma)$ for the full parameter vector. 
It may also be expressed as 
$$D(y,\theta)
  =(\hatt\beta-\beta)^\tr\Omega_n(\hatt\beta-\beta)/\sigma^2
  =\|\hatt\mu-\mu\|^2/\sigma^2=\sumin(\hatt\mu_i-\mu_i)^2/\sigma^2 \eqno(5.1)$$
in terms of the vector $\mu$ of means $\mu_i=x_i^\tr\beta$
and their fitted values $\hatt\mu_i=x_i^\tr\hatt\beta$. 

One may of course include other discrepancies too
for one's analysis, like 
$\maxin|y_i-x_i^\tr\beta|/\sigma$
or $\max_t|n^{-1}\sumin I\{(y_i-x_i^\tr\beta)/\sigma\le t\}-\Phi(t)|$,   
writing $\Phi$ for the cumulative standard normal. 
Theory and computational schemes may be developed
for these and other $D$ functions, following
the arguments and methods of this section. 
Any Bayesian regression analysis could in principle 
be supplemented with ppp and cppp analysis,
along the lines we give here. 

\subsection
{\sl 5.1. The case of known $\sigma$.}
Calculations for the ppp and cppp are rather easier 
and more immediately interpretable
for the case where the data standard deviation $\sigma$
is taken known, so we study that case first.
Assume $\beta$ has a prior distribution of the form
$\N_p(\beta_0,\sigma^2(c_0\Omega_0)^{-1})$, with 
a matrix $\Omega_0$ and a scalar $c_0$. The parameterisation
is a bit redundant, since $c_0$ may be taken into the
$\Omega_0$, but it is useful to start with the covariance
structure and then explore different levels of 
sharpness ($c_0$ large) or vagueness ($c_0$ small)
for the prior. One finds using multivariate normal 
theory and some matrix algebra that $\beta$ 
given data is normal, with
$$\eqalign{
\E(\beta\midd\data)
&=\tilda\beta=(c_0\Omega_0+\Omega_n)^{-1}
  (c_0\Omega_0\beta_0+\Omega_n\hatt\beta^\obs), \cr
\Var(\beta\midd\data)
&=\sigma^2(c_0\Omega_0+\Omega_n)^{-1}. \cr}$$
Also, 
$\tilda\beta-\hatt\beta^\obs
   =(c_0\Omega_0+\Omega_n)^{-1}c_0\Omega_0
   (\beta_0-\hatt\beta^\obs)$. 
Since $D(y^\rep,\theta)\sim\chi^2_p$, independent of 
$D(y^\obs,\theta)$, this leads to the formula
$$\ppp(y^\obs)=\Pr\{\chi^2_p\ge (U+f)^\tr\Omega_n(U+f)\}, $$
where $U\sim\N_p(0,(c_0\Omega_0+\Omega_n)^{-1})$ 
and 
$f=(c_0\Omega_0+\Omega_n)^{-1}c_0\Omega_0(\hatt\beta^\obs-\beta_0)/\sigma$.
Computation would most often be simplest via simulation
of $U$ vectors. 
We note that when $c_0$ is large, $U$ goes to zero
and $f$ goes to $(\hatt\beta^\obs-\beta_0)/\sigma$,
making $\ppp(y^\obs)\arr p^*(y)$, the traditional 
p-value for testing $\beta=\beta_0$ using 
the $(\hatt\beta-\beta_0)^\tr\Omega_n(\hatt\beta-\beta_0)/\sigma^2$
test. On the other hand, when $c_0$ becomes small,
$f\arr0$ and $\ppp(y^\obs)\arr\half$ for each data set $y^\obs$. 

To calibrate the ppp we need to compute a number $B$ of 
$\ppp(y_k)$ values from simulated data sets $y_k$,
as per (4.5). We would draw these sets by first 
drawing $\beta_k$ from the posterior distribution
and then creating $y_k=X\beta_k+\eps_k$, i.e.~without
changing or resampling the covariate vectors. 
For each of these $B$ sets one would find the $\ppp(y_k)$
number via simulation of $A$ random vectors $U$, i.e.
$$\ppp(y_k)\doteq{1\over A}\sum_{j=1}^A
   \Pr\{\chi^2_p\ge (U_{k,j}-f(y_k))^\tr\Omega_n(U_{k,j}-f(y_k))\}. $$

The only simple case is that of $\Omega_0$ being proportional 
to $\Omega_n$. We may in that case without loss of generality 
take $\Omega_0=n^{-1}\Omega_n$, with appropriate 
`prior sample size' interpretation for the flexible factor $c_0$ 
in $\Var\,\beta=\sigma^2(n/c_0)\Omega_n^{-1}$; 
the $\Omega_n$ matrix indeed grows linearly with sample size $n$. 
We then have 
$$\tilda\beta={c_0\over c_0+n}\beta_0
  +{n\over c_0+n}\hatt\beta^\obs, \quad
  U\sim\N_p\Bigl(0,{n\over c_0+n}\Omega_n^{-1}\Bigr), \quad
  f={c_0\over c_0+n}{\hatt\beta^\obs-\beta_0\over \sigma}. $$
This leads to a representation for $(U+f)^\tr\Omega_n(U+f)$,
in terms of $W\sim\N_p(0,\Omega_n^{-1})$, as 
$$\eqalign{
{n\over c_0+n}&\Bigl(W+\Bigl({c_0+n\over n}\Bigr)^{1/2}{c_0\over c_0+n}
   {\hatt\beta^\obs-\beta_0\over \sigma}\Bigr)^\tr\Omega_n
   \Bigl(W+\Bigl({c_0+n\over n}\Bigr)^{1/2}{c_0\over c_0+n}
   {\hatt\beta^\obs-\beta_0\over \sigma}\Bigr) \cr
&\sim{n\over c_0+n}\chi^2_p\Bigl({c_0^2\over n}{1\over c_0+n}
   {(\hatt\beta^\obs-\beta_0)^\tr\Omega_n(\hatt\beta^\obs-\beta_0)
   \over \sigma^2}\Bigr). \cr}$$
In consequence, under the $\Omega_0=n^{-1}\Omega_n$ scenario, 
$$\ppp(y^\obs)=F_{p,p}\Bigl(1+{c_0\over n},
   {c_0^2\over c_0+n}{\kappa_n^\obs \over \sigma^2}\Bigr), \eqno(5.2)$$
in terms of 
$\kappa_n^\obs=n^{-1}(\hatt\beta^\obs-\beta_0)^\tr
  \Omega_n(\hatt\beta^\obs-\beta_0)$
and the non-central $F$ distribution function.  
\eject 


This also leads to an explicit formula for the cppp,
as follows:
$$\eqalign{
\cppp(y^\obs)
&=\Pr\{\ppp(Y)\le \ppp(y^\obs)\} \cr
&=\Pr\{(\hatt\beta(Y)-\beta_0)^\tr\Omega_n(\hatt\beta(Y)-\beta_0)/\sigma^2
   \ge (\hatt\beta^\obs-\beta_0)^\tr\Omega_n
   (\hatt\beta^\obs-\beta_0)/\sigma^2\} \cr
&=\Pr\{\chi^2_p\ge{c_0\over c_0+n}
  (\hatt\beta^\obs-\beta_0)^\tr\Omega_n
  (\hatt\beta^\obs-\beta_0)/\sigma^2\}. \cr}$$
We use here that under prior and model conditions, 
$\hatt\beta(Y)$ is normal with mean $\beta_0$
and variance $\sigma^2(1+n/c_0)\Omega_n^{-1}$. 

\subsection
{\sl 5.2. Full analysis with unknown $\sigma$.}
The traditional conjugate prior in this model 
takes an inverse gamma for $\sigma^2$ and 
a normal for $\beta$ given $\sigma$. In generalisation 
of material of Section~2.3, agree now to say that 
$(\lambda,\mu)=(1/\sigma^2,\beta)$
comes from the $\GN_p(\half a_0,\half b_0,\beta_0,c_0\Omega_0)$ 
distribution if
$$\lambda\sim\Gam(\half a_0,\half b_0)
  \quad {\rm and} \quad 
  (\beta\midd\lambda)\sim\N_p(\beta_0,
  \lambda^{-1}(c_0\Omega_0)^{-1}). \eqno(5.3)$$
Thus $\beta$ has prior mean $\beta_0$ and prior variance
$(\E\lambda^{-1})c_0^{-1}\Omega_0^{-1}$. 
We shall see that the posterior takes the form 
$$\{(\lambda,\beta)\midd\data\}\sim
  \GN_p\bigl(\half(a_0+n),\half(b_0+Q_0^\obs
   +(\hatt\beta^\obs-\beta_0)^\tr 
  K(\hatt\beta^\obs-\beta_0)),
  \tilda\beta,c_0\Omega_0+\Omega_n\bigr), \eqno(5.4)$$
in which 
$Q_0^\obs=\sumin(y_i-x_i^\tr\hatt\beta^\obs)^2=\|y-\hatt\mu^\obs\|^2$ 
is the least sum of squares and 
$$\tilda\beta=(c_0\Omega_0+\Omega_n)^{-1}
  (c_0\Omega_0\beta_0+\Omega_n\hatt\beta^\obs)
  \quad {\rm and} \quad 
  K=(c_0^{-1}\Omega_0^{-1}+\Omega_n^{-1})^{-1}. $$
Proving this is accomplished partly via 
a little algebraic lemma, which states that for 
vectors $a_0,a_1$ and positive definite matrices 
$G_0,G_1$, the following identity holds: 
$$(x-a_0)^\tr G_0(x-a_0)+(x-a_1)^\tr G_1(x-a_1)
   =(x-\tilda a)^\tr(G_0+G_1)(x-\tilda a)
    +(a_1-a_0)^\tr M(a_1-a_0), $$
where $\tilda a=(G_0+G_1)^{-1}(G_0a_0+G_1a_1)$ and 
$M=(G_0^{-1}+G_1^{-1})^{-1}$. 
To prove (5.4), start with the prior density for $(\lambda,\beta)$, 
which is proportional to  
$$\lambda^{a/2-1}\exp(-\half b\lambda)
   \lambda^{p/2}\exp\{-\half c_0\lambda(\beta-\beta_0)^\tr\Omega_0
   (\beta-\beta_0)\}. $$  
The likelihood for data is proportional to
$\lambda^{n/2}\exp[-\half\lambda\{Q_0^\obs
   +(\hatt\beta^\obs-\beta)^\tr\Omega_n(\hatt\beta^\obs-\beta)\}]$.
The posterior is then seen to become proportional to
$$\eqalign{
\lambda^{(a+n)/2-1+p/2}
 &\exp[-\half\lambda\{b+Q_0^\obs+(\beta-\tilda\beta)^\tr
    (c_0\Omega_0+\Omega_n)^{-1}(\beta-\tilda\beta) \cr
 &\qquad\qquad 
    +(\hatt\beta^\obs-\beta_0)^\tr K(\hatt\beta^\obs-\beta_0)\}], \cr}$$
using the algebraic identity above. 
This is of the required form; 
see also Remark 2.3.

This leads to an accessible scheme for computing 
the ppp, under the (5.3) prior, as follows. Under model conditions, 
conditional on $(\beta,\sigma)$, one knows 
that $\hatt\beta$ is a $\N_p(\beta,\sigma^2\Omega_n^{-1})$,
allowing $D(y^\rep,\theta)$ a representation of the form 
$V^\tr\Omega_n V$, where $V\sim\N_p(0,\Omega_n^{-1})$; 
hence $D(y^\rep,\beta)$ is a $\chi^2_p$. 
To work with $D(y^\obs,\theta)$, we represent 
for the posterior distribution $\beta$ 
for given $\lambda$ as $\tilda\beta+\lambda^{-1/2}U$,
where $U\sim\N_p(0,(c_0\Omega_0+\Omega_n)^{-1})$ 
and independent of the $\chi^2_p$ distribution 
of $D(y^\rep,\theta)$. Thus 
$$\eqalign{D(y^\obs,\theta)
&=\lambda(\tilda\beta-\hatt\beta^\obs
  +\lambda^{-1/2}U)^\tr\Omega_n
   (\tilda\beta-\hatt\beta^\obs+\lambda^{-1/2}U) \cr
&=(\lambda^{1/2}(c_0\Omega_0+\Omega_n)^{-1}
  c_0\Omega_0(\hatt\beta^\obs-\beta_0)+U)^\tr\Omega_n \cr
&\qquad (\lambda^{1/2}(c_0\Omega_0+\Omega_n)^{-1}c_0\Omega_0
   (\hatt\beta^\obs-\beta_0)+U). \cr}$$
It is now relatively easy to simulate 
a large number $A$ of $(\lambda_j,U_j)$ replicates, with 
$$\lambda\sim\Gam(\half a_n,\half b_n), \quad
  {\rm where\ }
  a_n=a_0+n {\rm\ and\ } 
  b_n=b_0+Q_0^\obs
  +(\hatt\beta^\obs-\beta_0)^\tr K(\hatt\beta^\obs-\beta_0). $$ 
This gives replicates of $D(y^\obs,\theta_j)$ and the 
required simulation approximation
$$\ppp(y^\obs)\doteq{1\over A}\sum_{j=1}^A
   \Pr\{\chi^2_p\ge D(y^\obs,\theta_j)\}. \eqno(5.5)$$
This is a more precise estimate than the general-recipe
version used in (1.3); the (5.5) option is available
here since $D(y^\rep,\theta)$ has the known $\chi^2_p$
distribution, regardless of $\theta$.

Formula (5.5) also lends itself to the computation 
of the cppp, through a double simulation regime
that computes $\ppp(y_k)$ for many simulated 
data set, as per the general (4.5) strategy. 

\subsection
{\sl 5.3. The case of proportional $\Omega_0$ and $\Omega_n$.}
Important simplifications are found  
for the special case where the prior variance 
of $\beta$ is specified as being proportional to 
the sample variance of its least squares estimator. 
In that case we may again take $\Omega_0=n^{-1}\Omega_n$, 
as above, thereby also giving a more precise interpretation 
of $c_0$ in relation to sample size. With efforts 
similar to those exuded in the previous subsection one finds 
$$D(y^\obs,\theta)={n\over c_0+n}
   \chi^2_p\Bigl(\lambda{c_0^2\over c_0+n}
   {1\over n}(\hatt\beta^\obs-\beta_0)^\tr\Omega_n
   (\hatt\beta^\obs-\beta_0)\Bigr), $$
leading to the formula 
$$\eqalign{
\ppp(y^\obs)
&=\Pr\Bigl\{\chi^2_p\ge{n\over c_0+n}
   \chi^2_p\Bigl(\lambda{c_0^2\over c_0+n}\kappa_n^\obs\Bigr)\Bigr\} \cr
&=\int_0^\infty F_{p,p}\Bigl({c_0+n\over n},
   \lambda{c_0^2\over c_0+n}\kappa_n^\obs\Bigr)
   g_n(\lambda)\,\d\lambda, \cr} \eqno(5.6)$$ 
in terms of 
$\kappa_n^\obs=n^{-1}(\hatt\beta^\obs-\beta_0)^\tr
   \Omega_n(\hatt\beta^\obs-\beta_0)$
and the posterior gamma density $g_n$ for $\lambda$
implicit in (5.4). Its parameters are now $a_n=a_0+n$ and 
$b_n=b_0+Q_0^\obs+c_0n\kappa_n^\obs/(c_0+n)$. 
The (5.6) result is a generalisation of (2.4). 

For cppp analysis it is not a difficult task 
to compute $\ppp(y_k)$ as in (5.6) for a high number 
$B$ of simulated data sets, leading by recipe (4.5) 
to the appropriate $\cppp(y^\obs)$ number. 
Special properties of the linear regression model
make it possible to simplify this step, however,
as we now demonstrate. 

{\smallskip\sl
{\csc Proposition.} 
Suppose the prior for $(\sigma,\beta)$ is as in (5.3),
with $\Omega_0=n^{-1}\Omega_n$, and that $Y$ given
these parameters really follows the linear regression
model $\N_n(X\beta,\sigma^2I_n)$. 
Letting $Z\sim\chi^2_p$, the ppp distribution is 
$$G(u)=\Pr\{\ppp(Y)\le u\}
   =\Pr\{F_{p,p}(1+c_0/n,(c_0/n)Z)\le u\}. $$
\smallskip}

This makes it easy to compute the necessary 
$\cppp(y^\obs)=G(\ppp(y^\obs))$ as 
the relative frequency of 
$F_{p,p}(1+c_0/n,(c_0/n)Z_k)\le\ppp(y^\obs)$ events, 
across a million copies $Z_k$ from the $\chi^2_p$, 
since the non-central $F_{p,p}$ function is implemented
in software packages like {\tt R};  
in particular there is no need to carry out the 
updating of the inverse gamma parameters 
or to perform the numerical integration in (5.6) each time. 
The $G$ distribution is close to a uniform for large $c_0$
(corresponding almost to a classic p-value for
testing $\beta=\beta_0$), is very tightly concentrated
around $\half$ for small $c_0$ (corresponding to
a nearly non-informative prior), with a sharp upper 
bound at $F_{p,p}(1+c_0/n,0)$. The case study
presented in the following section has $c_0=6.25$,
an intermediate case. 

\smallskip
{\csc Proof.}
Let $H=X(X^\tr X)^{-1}X^\tr=X\Omega_n^{-1}X^\tr$ be the 
familiar `hat matrix'. It is symmetric and idempotent,
and a standard result is that for given $(\sigma,\beta)$,  
$$X\hatt\beta=HY\sim\N_n(X\beta,\sigma^2H),
  \quad
  \hatt\eps=Y-X\hatt\beta=(I-H)Y\sim\N_n(0,\sigma^2(I-H)), $$
and that these two $n$-vectors are independent. 
Taking (5.3) into account for the distribution 
of $\beta$ given $\sigma$, one finds that the
vectors remain independent, for given $\sigma$, with 
$$X\hatt\beta\sim\N_n(X\beta_0,\sigma^2(1+n/c_0)H),
  \quad
  \hatt\eps\sim\N_n(0,\sigma^2(I-H)). $$
This implies $Q_0(Y)=\|\hatt\eps\|^2\sim\sigma^2(V^*)^\tr V^*$ and 
$$\kappa_n(Y)=n^{-1}(X\hatt\beta-X\beta_0)^\tr(X\hatt\beta-X\beta_0)
   =n^{-1}\sigma^2(1+n/c_0)V^\tr V, $$
in terms of $V^*\sim\N_n(0,I-H)$ and $V\sim\N_n(0,H)$,
with these two being independent. Hence
$\kappa_n(Y)=\sigma^2(n^{-1}+c_0^{-1})Z$
and $Q_0(Y)=\sigma^2Z^*$, where $Z$ and $Z^*$ 
are $\chi^2$ distributed and independent with 
$p$ and $n-p$ degrees of freedom. This simplifies
the updated parameters of the Gamma distribution
for $\lambda=1/\sigma^2$, to $a_n=a_0+n$
and $b_n=b_0+W_n$ where $W_n\sim\sigma^2(Z+Z^*)$, and,
more importantly, via (5.6), 
$$\ppp(Y)=\Pr\Bigl\{\chi^2_p\ge{n\over c_0+n}
   \chi^2_p\Bigl({1\over \sigma^2}{c_0^2\over c_0+n}
   \sigma^2\Bigl({1\over n}+{1\over c_0}\Bigr)Z\Bigr)\Bigr\}, $$
in terms of a central and a non-central $\chi^2_p$ 
variable that both are independent of $Z$. 
The simplifying point is that the $\sigma^2$ factors cancel out.
The claim follows. 
\square

\smallskip
{\csc Remark 5.1.}
When sample size $n$ is large, $n^{-1}\Omega_n$
will be close to a limiting covariance matrix
$\Sigma$ for the covariates, and $\hatt\beta^\obs$
will be close to the true regression coefficient
vector $\beta_\true$. Some analysis shows that
$\cppp(y^\obs)\arr\Pr\{\chi^2_p\ge\conf^2\}$, where 
$$\conf=c_0^{1/2}\{(\beta_\true-\beta_0)^\tr\Sigma
   (\beta_\true-\beta_0)\}^{1/2}/\sigma $$
acts as a conflict measure between prior and model,
also dependent on prior sample size $c_0$. This
generalises Remark 4.1. See also Section 10.1.
\square

\section
\centerline{\bf 6. Case study: regressing speedskaters}

\hop
In the annual World Sprint Championships in speedskating,
competitors skate the 500 m and 1000 m distances 
on Saturday, and then the same distances on Sunday. 
The champion is the skater with the lowest combined
score, defined as $t_1+t_2/2+t_3+t_4/2$, where 
$t_1,t_2,t_3,t_4$ are the results (in seconds) 
of the skater for the four distances; see Table 1, 
which gives the top of the result list 
for the 2005 championships, held at the Olympic Oval 
in Salt Lake City (with 33 participants from 12 countries).
In this section we will illustrate 
the general methods of Section 5 for the regression model 
that pertains to predicting and understanding 
the result of the fourth distance in terms of times 
achieved for the previous three distances. 
In this case there would be substantial prior
knowledge (along with excitement and speculations), 
among skaters and the millions of television viewers, 
relating to the parameters and the fruitfulness 
of the model, making Bayesian analysis 
a relevant and interesting enterprise.  

\def\hd{\hskip10pt}
\def\he{\phantom{1}}

{{\medskip
\baselineskip11pt
\settabs
\+\quad 10    &Erb Erben Wennemars  &140.755\hd       
   &35.70 (9)\hd &1:09.46 (1)\hd &35.63 (7)\hd &1:09.39 (1) \cr
\+\quad {\he}1     &Erben Wennemars  &137.310   
   &34.75 (4)  &1:08.30 (4)  &34.68 (3)  &1:07.46 (1) \cr
\+\quad {\he}2 &Jeremy Wotherspoon   &137.820  
   &34.65 (3)  &1:08.40 (5)  &34.67 (1)  &1:08.60 (8) \cr
\+\quad {\he}3 &Joey Cheek           &137.975 
   &34.88 (8)  &1:08.59 (7)  &34.70 (4)  &1:08.20 (4) \cr        
\+\quad {\he}4 &Masaaki Kobayashi    &138.050 
   &34.76 (5)  &1:08.28 (3)  &34.90 (7)  &1:08.50 (5) \cr
\+\quad {\he}5 &Dmitri Lobkov        &138.100 
   &34.63 (2)  &1:08.78 (9)  &34.67 (1)  &1:08.82 (10) \cr       
\+\quad {\he}6 &Jan Bos              &138.470 
   &35.25 (14) &1:08.00 (1)  &35.26 (11) &1:07.92 (3) \cr        
\+\quad {\he}7 &Shani Davis          &138.715 
   &35.43 (20) &1:08.04 (2)  &35.43 (17) &1:07.67 (2) \cr        
\+\quad {\he}8  &Casey FitzRandolph  &138.770 
   &34.99 (9)  &1:08.84 (11) &35.11 (10) &1:08.50 (5) \cr
\+\quad {\he}9  &Joji Kato           &139.000 
   &34.80 (6)  &1:09.54 (17) &34.89 (6)  &1:09.08 (14) \cr       
\+\quad 10 &Hiroyasu Shimizu         &139.230 
   &35.12 (10) &1:09.49 (16) &34.82 (5)  &1:09.09 (15) \cr 
\+\quad &. . . \cr
\smallskip\narrower\noindent\sl
{\csc Table 1.} 
The best ten skaters in the 2005 World Sprint Championships,
held in Salt Lake City, USA, January 22--23. 
The results are the combined point sum,
then the 500 m and 1000 m results for Saturday,
followed by the 500 m and 1000 m results for Sunday
(with numbers in parentheses indicating ranking).
\smallskip}}

\subsection
{\sl 6.1. Setting the prior.}
The natural model to consider takes
$$y_i=b_0+b_1x_{i,1}+\cdots+b_px_{i,p}+\eps_i
   =b_0+x_i^\tr b+\eps_i
   \quad {\rm for\ }i=1,\ldots,n, $$ 
in terms of a $p$-dimensional covariate vector $x_i$ 
for each of $n$ individuals, where the $\eps_i$s 
are independent zero-mean normals with standard 
deviation $\sigma$. For the speedskating applications,
$y_i$ is the 1000 m Sunday result, while 
$x_{i,1},x_{i,2},x_{i,3}$ are the results 
of the three previous distances. Our model 
is intended to convey the main mechanism of
achievements during the World Championships 
among the say 25--30 best skaters of the world, 
during races without falls or accidents. 
Such `negative outliers' are therefore screened out 
and not allowed to enter the model, while 
on the other hand unusually splendid results,
like a new world record, are certainly allowed. 
We shall test this model with ppp-scrutiny 
associated with discrepancy measure (5.1). 

Setting a clear prior for $(b_0,b,\sigma)$ is 
not an easy exercise. It is not quite sufficient  
to have $(b_1,b_2,b_3)$ centred at $(0,1,0)$, 
even though $x_{i,2}$ is a reasonable guess 
for $y_i$; this viewpoint does not take into 
account that $y_i$ also tends to be positively 
associated with both $x_{i,1}$ and $x_{i,3}$. 
One option here would be to use last year's 
tables of results to provide that competition's 
posterior distribution of parameters, as 
a prior for this year's model, perhaps scaled 
down in precision so as to not be too informative. 
To make the exercise more realistic, however, 
having likely future applications of our theory 
in mind, we adopt the attitude that we should 
construct our prior from subjective (but 
substantiated) beliefs about the parameters. 
We argue that the prior knowledge in this situation
(and, we suggest, in many other) is most easily quantified 
in terms of (i) the overall level and its variability 
(expected average and standard deviation for the $y_i$s) 
and (ii) the correlations between covariates 
$x_{\cdot,j}$ and the outcome $y$. We therefore develop 
a method for bringing such prior knowledge 
into a proper prior for the model parameters. 

It is helpful first to centre the covariates 
by subtracting the averages $\bar x_{\cdot,j}$ 
from the $x_{i,j}$s above. Considering this done, 
$\bar y=b_0+\bar\eps$, giving $b_0$ a clear 
interpretation as the expected average result.
Furthermore, 
$n^{-1}\sumin y_ix_i=S_nb+n^{-1}\sumin x_i\eps_i$,
where $S_n=n^{-1}\sumin x_ix_i^\tr$ is the 
empirical variance matrix of the covariate vectors, 
with elements say $s_{j,k}$ for $j,k=1,\ldots,p$. 
While the considerations that follow actually might
invite also other types of priors we wish for this
illustration to follow the development and recipes 
of Section 5, so we are to specify a prior for $\sigma$ 
and for the regression curve parameters given $\sigma$ 
of the form
$$\pmatrix{b_0 \cr b \cr}\midd\sigma
   \sim\N_{p+1}(\pmatrix{\bar b_0 \cr \bar b \cr},
   \sigma^2{1\over c_0}\pmatrix{1 &0 \cr 0 &S_n^{-1} \cr}). \eqno(6.1)$$
Since $n^{-1}\sumin(y_i-\bar y)^2$
is seen to have mean value $b^\tr S_nb+\sigma^2(1-1/n)$,
which we approximate with $b^\tr S_nb+\sigma^2$,  
we see that the correlations 
$\rho=(\rho_1,\ldots,\rho_p)^\tr$
between $y$ and the respective covariates 
$x_{\cdot,j}$ may be represented as 
$$\rho_j={(S_nb)_j\over (b^\tr S_nb+\sigma^2)^{1/2}s_j}
  \quad {\rm for\ }j=1,\ldots,p, \eqno(6.2)$$
where $s_j=s_{j,j}^{1/2}$ is the standard deviation 
of $x_{\cdot,j}$. We therefore need a mechanism 
for turning prior information about $\rho$ 
into a prior for $b$. We assume that the list 
of covariate vectors is available to aid us
in fine-tuning the prior, as with the speedskating data. 
 
To work with this, we solve the (6.2) equations for $b$.
They may be expressed as 
$S_nb/(b^\tr S_nb+\sigma^2)^{1/2}=v=D_n^{1/2}\rho$, 
where $D_n=\diag(s_1^2,\ldots,s_p^2)$ 
is the matrix with empirical covariate variances 
down its diagonal; thus $v$ has components $s_j\rho_j$. 
The solution is 
$b=\sigma S_n^{-1}v/(1-v^\tr S_n^{-1}v)^{1/2}$. 
To follow the intended scheme we take 
$$S_n^{-1/2}v/(1-v^\tr S_n^{-1}v)^{1/2}
   =z_0/\sigma+\tau N\sim\N_p(z_0/\sigma,\tau^2I_p), $$
for a suitable location $z_0$ and scale parameter $\tau$,
in terms of an $N\sim\N_p(0,I_p)$. This also means that 
$$(b\midd\sigma)\sim\N_p(S_n^{-1/2}z_0,\sigma^2\tau^2S_n^{-1}), \eqno(6.3)$$
which is consistent with the conjugate set-up of (5.3) and (6.1).
To fine-tune the parameters, 
solve for $v=(s_1\rho_1,\ldots,s_p\rho_p)^\tr$ to find
$$S_n^{-1/2}v={z_0/\sigma+\tau N
   \over (1+\|z_0/\sigma+\tau N\|^2)^{1/2}}. \eqno(6.4)$$
Our strategy is as follows: 
(i) Put up prior guess parameters $\rho_{j,0}$
for the correlations $\rho_1,\ldots,\rho_p$. 
This must in particular be done in a manner 
which reflects $v^\tr S_n^{-1}v<1$, for $v=D_n^{1/2}\rho$.  
(ii) Then, for each of a sequence of trial values
of the parameter $\tau$, solve the $p$ equations
$$g(z_0)=\E{z_0/\sigma_0+\tau N\over (1+\|z_0/\sigma_0+\tau N\|^2)^{1/2}}
   =S_n^{-1/2}v_0=S_n^{-1/2}D_n^{1/2}\rho_0 \eqno(6.5)$$
for $z_0$, utilising also an initial prior guess
value $1/\sigma_0$ for $1/\sigma$. In practice 
we solve these equations using stochastic 
simulation to compute $g_1(z_0),\ldots,g_p(z_0)$, 
since no formulae are available, and a non-linear 
minimisation algorithm like {\tt nlm} in the {\tt R} 
software package to minimise 
$\sum_{j=1}^p\{g_j(z_0)-(S_n^{-1/2}v_0)_j\}^2$.
(iii) For the trial value of $\tau$, and the found 
$z_0$, one generates $\rho$ vectors from $\rho=D_n^{-1/2}v$,
where $v$ is taken according to (6.4); in particular
one may monitor the spread of the $\rho_j$s via
their standard deviations and correlations (again,
there are no explicit formulae to help us, hence
the simulations). 
(iv) The procedure stops when a value of $\tau$ 
is found that well reflects the uncertainty level for the
$\rho_j$s around the prior guesses $\rho_{j,0}$.

The procedure would perhaps have looked more precise
and elegant were it possible to produce a clear 
prior covariance matrix, say $\Lambda$, for $\rho$,
and then to choose $\tau$ to match the mean 
of $v^\tr S_n^{-1}v$, which is 
$v_0^\tr S_n^{-1}v_0+\Tr(D_n^{1/2}\Lambda S_n^{-1/2})$,
with the mean of $\|z_0/\sigma+\tau N\|^2/(1+\|z_0/\sigma+\tau N\|^2)$,
from (6.4). The problem with this is the difficulty of
setting a good $\Lambda$ matrix, in that attempts 
at doing this would clash with the structure imposed by
(6.4). Our strategy avoids this quandary.

\subsection
{\sl 6.2. ppp and cppp analysis of the 2005 speedskating model.}
We carried out the above exercise for the World Championships
2005 data. A total of six skaters were eliminated from 
the analysis, after falls, disqualifications, and 
injuries causing some skaters to not finish all four
distances. This left 27 healthy skaters in the data matrix. 
We used prior guess values .6, .8, .6
for the correlations $(\rho_1,\rho_2,\rho_3)$ 
of $x_{\cdot,1},x_{\cdot,2},x_{\cdot,3}$
with $y$, and $1/0.75$ for $1/\sigma_0$; 
these values were elicited based on discussion amongst
the authors and speedskating compatriots, 
and are meant to be based on solid experiences;
for more background, see e.g.~Hjort (2003a, 2005). One had  
$$S_n=\pmatrix{0.259 &0.252 &0.196 \cr
               0.252 &1.209 &0.223 \cr
               0.196 &0.223 &0.239 \cr}, $$
and we could for each trial value $\tau$ solve (6.5)
for $z_0$, in terms of 
$S_n^{-1/2}D_n^{1/2}\rho_0=(.094,\allowbreak .756,.294)^\tr$. 
Monitoring spread and correlations 
in the prior distribution for $\rho$ we settled 
on $\tau=.4$ to reflect prior beliefs. This corresponds 
to `prior sample size' $c_0=1/\tau^2=6.25$, 
to standard deviations $.132, .084, .133$ for the 
three correlations, centred around $.6,.8,.6$. 
The $(\rho_1,\rho_2,\rho_3)$ distribution is far from normal,
in its constrained space determined by 
$\rho^\tr D_n^{1/2}S_n^{-1}D_n\rho<1$.
The correlation correlations are $-.444, .634, -.521$ 
for $(\rho_1,\rho_2)$, $(\rho_1,\rho_3)$, and $(\rho_2,\rho_3)$.
We have also $S_n^{-1/2}z_0=(0.398,1.018,0.856)^\tr$ in (6.3),
i.e.~the prior mean $\bar b$ in (6.1). As 
prior mean $\bar b_0$ for $b_0$ we use the mean
of all Saturday's 1000 m results, which is 1:09.46. 

It remains only to specify suitable parameters 
$(\half a_0,\half b_0)$ for the gamma prior for 
$\lambda=1/\sigma^2$, to keep with the (5.3) recipe. 
To this end we may first show that 
$$b^\tr S_nb=\sigma^2{v^\tr S_n^{-1}v\over 1-v^\tr S_n^{-1}v}
  \quad {\rm and} \quad 
  \kappa^2=\Var\,y_i={\sigma^2\over 1-v^\tr S_n^{-1}v}, $$
which leads to the identity 
$\sigma=\kappa(1-v^\tr S_n^{-1}v)^{1/2}$
in terms of the standard deviation $\kappa$ 
for the $y_i$ results themselves. But for $\kappa$
we have quite relevant prior information after 
watching Saturday's 1000 m results, with standard deviation 
$\hatt\kappa=1.121$. We therefore select $(\half a_0,\half b_0)$
such that first 
$$\E{1\over \sigma^2}={a_0\over b_0}
  {\rm\ matches\ }
   {1\over (\sigma^*)^2}={1\over \hatt\kappa^2}
   {1\over 1-v^\tr S_n^{-1}v}={1\over 0.580^2}, $$
and, secondly, such that $\Var(1/\sigma^2)=2a_0/b_0^2$ 
matches the estimated variance 
$$\Var\Bigl({1\over \hatt\kappa^2}{1\over 1-v^\tr S_n^{-1}v}\Bigr)
   ={1\over \kappa^4}{2m^2\over (m-2)^2(m-4)}
   {1\over (1-v^\tr S_n^{-1}v)^2}
   \doteq{1\over 0.580^4}{2m^2\over (m-2)^2(m-4)}. $$
Here we use $\hatt\kappa^2\sim\kappa^2\chi^2_m/m$ with 
$m=n-1$. The procedure yields $(a_0,b_0)=(18.746,6.307)$. 

After the hard work of setting a proper prior 
it is a pleasure to finally watch Sunday's 1000 m races, 
an easy task to produce the ppp and cppp numbers, 
following the results of Section 5, and to update
our prior for the correlation and regression parameters.
The correlation uncertainty updating is of interest 
for the actual application (the posterior is e.g.~centred
at about $(.45,.79,.42)$), but is uncorrelated 
with the main story of this article, so we report instead
on the ppp analysis. 
Formula (5.6) gives $\ppp(y^\obs)=.444$. 
Applying the Proposition of Section 5 one 
can check, via a million simulations, that 
the $\ppp(Y)$ distribution in this case has 
90\% of its probability mass on $[.404,.563]$, 
for example, so .444 is closer to actual surprise
than what an ordinary (but here naive) p-value interpretation 
would suggest. One finds $\cppp(y^\obs)=.135$. 
We may conclude that there is no conflict
between our carefully constructed prior and the data,
as monitored via the canonical discrepancy function
$D(y,\theta)=\sumin(\hatt\mu_i-\mu_i)^2/\sigma^2$.  
It might be of interest to try other discrepancy
functions, checking adequacy of other aspects of 
prior and model, like asymmetry, but we abstain from
doing so here. 

\section
\centerline{\bf 7. ppp and cppp analysis of two bird survival models}

\hop
Brooks, Catchpole and Morgan (2000; henceforth BCM) 
analysed recapture data for the European Dipper species 
{\sl (Cinclus cinclus)} using ppp values. The same data 
have previously been analysed by 
Lebreton, Burnham, Clobert and Anderson (1992).
They are in the form of a triangular $6\times6$ array 
in conjunction with a vector giving the number of released 
individuals for the six years in question; see below. 
Here we offer a re-analysis of their data, utilising
our cppp tools, and reach conclusions partly different 
from those of BCM regarding adequacy of suggested models. 

\subsection
{\sl 7.1. The data, the model, and the discrepancy.}
The data are given in Table 2. The models considered by BCM 
involve up to twelve parameters: for $i=1,\ldots,6$, 
$\phi_i$ represents the probability that a given bird 
survives year $1980+i$, while $p_i$ represents the 
probability of capturing a particular bird in year $1980+i$.

\def\ha{\quad}
\def\hb{\hskip10pt}

{\smallskip\obeylines\baselineskip12pt
\settabs\+\indent&Release year \quad&Released \quad&Recaptured: \quad
&1982 &1983 &1984 &1985 &1986 &1987 \cr
\+&Release year &Released &Recaptured: &1982 &1983 &1984 &1985 &1986 &1987 \cr
\+\ha &\hb 1981   &\hb 22     &&~11  &~~2  &~~0 &~~0  &~~0  &~~0 \cr
\+\ha &\hb 1982   &\hb 60     &&~~-  &~24  &~~1 &~~0  &~~0  &~~0 \cr
\+\ha &\hb 1983   &\hb 78     &&~~-  &~~-  &~34 &~~2  &~~0  &~~0 \cr
\+\ha &\hb 1984   &\hb 80     &&~~-  &~~-  &~~- &~45  &~~1  &~~2 \cr
\+\ha &\hb 1985   &\hb 88     &&~~-  &~~-  &~~- &~~-  &~51  &~~0 \cr
\+\ha &\hb 1986   &\hb 98     &&~~-  &~~-  &~~- &~~-  &~~-  &~52 \cr
\smallskip}
{\smallskip\narrower\noindent\baselineskip12pt
{\csc Table 2.} 
{\sl Recapture data for the European Dipper.}
\smallskip}

\noindent
The multiplicative multinomial model assumption 
can for such recapture data be shown to lead 
to a likelihood of the form
$$L(\phi,p)={\rm const.}\,\Delta(\phi,p)
   \prod_{i=1}^6\prod_{j=i+1}^6
   \Bigl\{\phi_i p_j\prod_{k=i+1}^{j-1}
   \phi_k(1-p_k)\Bigr\}^{y_{i,j}}, $$ 
where $\Delta$ is the contribution to the likelihood 
for the never-captured birds, and where `empty products' 
are equal to 1; see BCM for details. The 
discrepancy measure used in BCM is 
$$D(y,(\phi,p))=\sum_{i,j}(y_{i,j}^{1/2} 
   -e_{i,j}^{1/2})^2, \quad
  {\rm where\ }e_{i,j}=\E_{\rm model}\,y_{i,j}. $$
We have analysed two versions of this general model.
The large model (T/T in the terminology of BCM) 
employs all twelve parameters $(\phi_i,p_i)$ 
while the small model (C/C in their terminology) 
takes all $\phi_j$s equal and all $p_j$s equal, 
thus having two parameters. 

For both cases we choose to use independent and 
uniform priors for the parameters in question. 
We do this for two reasons. Firstly, the more 
focussed priors also briefly worked with by BCM 
do not appear to fit data particularly well, 
and BCM give little indication that they 
represent actual prior knowledge of the parameters.
Secondly, they quote the posterior means and 
standard deviations only for the uniform prior. 
It turns out that our computed ppp values 
are slightly, but statistically significantly, 
different from those quoted by BCM
(for the full model with uniform prior, 
we find a ppp value of .075, 
while BCM quote the number .086; 
similarly, we find ppp equal to .060 where 
BCM give .069, for the small model). 
By verifying the quoted posterior means and 
standard deviations, we have eliminated the MCMC 
simulation as a possible reason for these 
differences. Although we have not been able to 
reproduce the exact ppp values of BCM, the 
differences are small enough to be ignored 
in the present setting of general analysis. 

\subsection
{\sl 7.2. ppp and cppp analysis of the two models.}

{\sl The large model.}  
We have sampled the distribution of ppp along the lines 
indicated above, and found $\ppp(y^\obs)=.075$. 
From plots (not shown here) of the 
sampled cumulative distribution function 
and estimated density for the ppp, 
based on 500 simulated values, we see that 
the distribution function is clearly S-shaped,
and the density clearly unimodal. 
The sample standard deviation is $.172$,  
compared to $.289$ for a uniformly distributed p-value. 
More important is the fact that only a single one
of our 500 simulated ppp values was below .075, giving
an estimated cppp value (the true surprise level) of .002. 



{\sl The small model.}
In this case we find $\ppp(y^\obs)=.060$.
Again we studied the sample cumulative distribution function 
and estimated density for the ppp. 
We found here that the ppp distribution is closer 
to a uniform distribution, compared to the twelve 
parameter case. The sample standard deviation 
of ppp is now $.257$. With calibration we 
find $\cppp=.022$ (11 out of 500 simulated $\ppp(Y)$ 
values were less than .060). We therefore 
conclude that the given data are much less 
likely under the large model than under the small 
one, although the nominal ppp values led
BCM to the opposite relation.


\section
\centerline{\bf 8. Comparing nonparametric with parametric models}

\hop
Our cppp can be interpreted as a `quantification of 
surprise', as monitored by the discrepancy function. 
It is a strength of this approach that it puts all 
surprises on the same footing, so to speak, 
namely the uniform scale on the unit interval.
In particular, there are no conceptual difficulties 
with comparing the cppp from a parametric model 
specification with that of a nonparametric one. 
We indicate how this might be worked with  
in two situations. 

\subsection
{\sl 8.1. Nonparametric vs.~parametric cdf.}
Consider independent data $y_1,\ldots,y_n$ 
drawn from a distribution $F$, where two quite 
different priors are under consideration for $F$. 
The first is to take $F(t)=\Phi((t-\mu)/\sigma)$ 
normal, with a prior $\pi(\mu,\sigma)$ on its 
parameters. The second does not bound $F$ to any
parametric description, and takes $F\sim\Dir(aF_0)$,
a Dirichlet process with centre distribution $F_0$
and concentration parameter $a$; see e.g.~Hjort (2003b)
for a review of nonparametric Bayesian statistics. 
For discrepancy measure we use 
$$D(y,F)=\rootn\|F_n-F\|=\rootn\max_t|F_n(t)-F(t)|, $$
in terms of the empirical distribution function $F_n$
of the data. The $\rootn$ factor is of no consequence 
for the actually computed ppp and cppp numbers, 
but is there to better understand the situation 
when $n$ grows. In fact, the 
$D(y^\rep,F)=\rootn\|F_n^\rep-F\|$
is then close to being the maximum absolute value of 
$W^0(F(t))$, where $W^0$ is a Brownian bridge, by
classic empirical process theory, see e.g.~Billingsley 
(1968, Ch.~4). In particular, 
$D(y^\rep,F)\arr_d \|W^0\|=\max_t|W^0(t)|$,
both when $F$ is fixed and when $F$ is selected by 
some posterior mechanism. 

\def\appr{{\rm appr}}

Let us now assume that the $y_i$ data really follow
some continuous $F_\true$ distribution. We are to study
the behaviour of the nonparametric and normal-parametric
prior specifications, say $\ppp_\Dir$ and $\ppp_\N$. 
First study the nonparametric prior. Here $F$ given data is 
a Dirichlet process with parameter $aF_0+nF_n^\obs$.
One may show that 
$\rootn(F-F_n^\obs)\arr_d \tilda W^0(F_\true(\cdot))$
a.s., where $\tilda W^0$ is another Brownian bridge,
independent of $W^0$; this follows e.g.~from work of
Hjort and Ongaro (2005). Hence
$$\ppp_\Dir(y^\obs)
  =\Pr\{\rootn\|F_n^\rep-F\|\ge \rootn\|F_n^\obs-F\|\midd\data\} 
  \arr\Pr\{\|W^0\|\ge \|\tilda W^0\|\}=\half $$
a.s., as $n$ grows towards infinity. Then consider the 
normal-parametric prior, for which $F$ given data is 
a random normal distribution function, with 
$(\mu,\sigma)$ drawn from the appropriate posterior 
density $\pi(\mu,\sigma\midd\data)$. Here 
$\|F_n^\obs-F\|$ goes a.s.~to $\|F_\true-F_\appr\|$, 
essentially by the Glivenko--Cantelli theorem, 
where $F_\appr(t)=\Phi((t-\mu_\true)/\sigma_\true)$ is 
the best parametric approximant to $F_\true(t)$,
inside the normal family. It follows that 
$\rootn\|F_n^\obs-F\|$ goes to infinity a.s., 
as long as the real $F_\true$ is not fully equal 
to a normal distribution function, and, in particular, 
$\ppp_\N(y^\obs)\arr0$ a.s.~as $n\arr\infty$.   

The results above help us understand the behaviour 
of the two ppp measures, for large $n$. We would 
also need to calibrate these, to reach the more 
interpretable cppp values given data. This can be 
done via double simulation, as per the general 
guidelines about this laid out in Section 4.2.

In some situations interest focusses more on 
some interest parameter $\kappa(F)$ than on the full
distribution $F$, say the interquartile range
or the skewness. In such cases one might prefer 
working with discrepancy measures of the type 
$D(y,F)=n\{\kappa(F_n)-\kappa(F)\}^2$. Here 
the parametric model could win, even when it is not
fully correct as such. Again, ppp and cppp analysis
may be carried out. 

\subsection
{\sl 8.2. Nonparametric vs.~parametric hazard rate models.}
Bayesian analysis of survival and event history data
appear to fall into one of two separate categories,
viz.~the parametric and the nonparametric. Only
rarely does one see any formal justification for 
choosing one path over the other. The cppp analysis
makes it in principle easy, conceptually and operationally,
to do such a comparison. The parametric model might 
use a Weibull cumulative hazard rate $H(t)=(\theta t)^\gamma$,
with a prior on $(\theta,\gamma)$, where the nonparametric
alternative could use a Beta process. As discrepancy
measure one might use 
$D(y,H)=\int_0^\tau w(t)\{H_n(t)-H(t)\}^2\,\d t$,
for a weight function $w$, involving the Nelson--Aalen
estimator $H_n$ for $H$ over a time observation window $[0,\tau]$.

\section
\centerline{\bf 9. ppp and cppp: facts and fiction} 

\hop
Various issues related to the rapidly evolving 
area of Bayesian model evaluation border on controversy,
and there are certainly conflicting views on the use
and limitations of ppp measures and various other 
simulation-based model or model-prior diagnostics. 
Below we attempt to clarify our own views on some
of these issues, in the light of our findings
and applications. 

\subsection
{\sl 9.1. The ppp needs a scale.}
A ppp number computed from data is a well-defined 
probability, but it has a difficult interpretation, 
due to the `double use of data' aspect of its
construction. In particular an observed $\ppp=.28$, say, 
cannot be taken as `support for' or 
`criticism against' the model, and neither can
it easily be compared with other ppp values. 
In some cases .28 can signal extreme surprise, 
in the sense of being a highly unlikely value 
under model conditions. This may also be true for 
a value of say .52, even when the mean value of 
the $\ppp$ distribution is .50. 
Thus an appropriate scale is called for 
in order to compute one's surprise level,
and this scale needs to take into account 
which data sets are considered likely or not. 
We consider the cppp of (1.4) a fairly canonical 
calibration, transforming the ppp to the uniform scale. 
Other transformations are sometimes relevant,
as discussed in Section 4.3. 

Let us for illustration go back to the speed of light
data described in the Example of Section 1, 
where the discrepancy 
$D=|y_{(61)}-\mu|-|\mu-y_{(6)}|$ is used for
assessing normality `in the middle of the distribution'.
We now examine the `real surprise level' of ppp 
scores under some natural priors, namely the 
inverse gamma-normal class defined in (2.2)
with parameters $(a_0,b_0,\mu_0,c_0)=(c/25,c,30,c)$.
This corresponds to prior guess 30 for $\mu$ 
and prior mean $a_0/b_0=1/25$ for $1/\sigma^2$, 
while $c$ in a reasonable sense is the prior sample size,
in view of the updating mechanism for $\GN$ priors
laid out in Section 2.3. In particular the vague
prior case corresponds to $c\arr0$. 
The table below shows $\ppp$ and $\cppp$ values
for different $c$ values, using the double simulation 
method to compute the cppp numbers, 
with $A=B=2000$ in (1.3) and (4.5): 
$$\matrix{
c   &\ppp  &\cppp & &c  &\ppp &\cppp \cr
\noalign{\smallskip}
.5  &.216  &.067  & &10  &.280 &.146 &\cr
1   &.218  &.069  & &20  &.333 &.227 &\cr
2   &.223  &.083  & &30  &.386 &.335 &\cr
3   &.228  &.089  & &40  &.422 &.396 &\cr
4   &.238  &.093  & &50  &.470 &.466 &\cr
5   &.249  &.107  & &100 &.588 &.619 &\cr} $$ 
The table indicates that $\ppp(y^\obs)=.208$ might be judged
to be on the statistical borderline of surprise after all
(despite the conclusion of Gelman et al.~(2004, p.~164)). 
This is also consistent with results from a couple
of frequentist checks of the type 
$T=(y_{(61)}-y_{(6)})/\hatt\sigma$, with any
robust scale estimate; normality is rejected at
level .03 if $\hatt\sigma$ is the mean-mad
estimator $n^{-1}\sumin|y_i-{\rm med}(y)|$, for example. 

\subsection
{\sl 9.2. One discrepancy's cppp is not another discrepancy's ppp.} 
One reaction we have met is that the cppp 
for a discrepancy $D(y,\theta)$ ought to be  
equivalent to the original ppp for another 
discrepancy $D^*(y,\theta)$. We shall however 
argue that this is typically not the case. 
In this sense the cppp is therefore a `new invention'.

The $\cppp$ is by construction distributed as a $U(0,1)$.
Hence the desired $\ppp^*$, corresponding to $D^*$, 
must also be $U(0,1)$. Define 
$h(z,\theta)=\Pr\{D^*(y,\theta)\ge D^*(z,\theta)\}$,
where $y\sim f(y,\theta)$. It follows from Theorem 1 of 
Meng (1994) that $\ppp^*\sim U(0,1)$ 
only if $h(y^\obs,\theta)$ is constant under the posterior 
distribution of $\theta$, for almost all $y^\obs$.
This can only happen if the posterior distribution of
$\theta$ does not affect the distribution of 
$y\midd\theta$ in ways detected by $D^*$. 
In particular, defining $D^*$ as a function of $y$ only 
is not sufficient, as $h(y^\obs,\theta)$ 
will still depend on its second parameter 
through the distribution of $y\midd\theta$. 
Essentially, $D^*$ can only monitor aspects of 
$(y,\theta)$ for which the prior distribution 
of $\theta$ is not random.
On the other hand, the desired $D^*$ must be able to
detect the same discrepancies as $D$, in order to achieve 
$\ppp^*=\cppp$. We conclude that such a $D^*$ is unlikely 
to exist, except in cases where the randomness 
of the prior distribution is irrelevant for the discrepancy $D$.

\subsection
{\sl 9.3. Some priors are more proper than others.}
There are different schools of thought 
among scientists that apply Bayesian methods;
Good (1983) reminds us that there are at least 46,656
different types of Bayesian thinking.  
Purists may even imagine that the parameters of 
Nature were drawn according to the prior distribution 
when the world was created. At the other extreme, 
many statisticians use Bayesian methods as a means 
of estimating parameters through posterior means 
of MCMC samples, partly because this appears to be
(at this stage) more computationally convenient 
than frequentistic estimation.
The latter group typically uses improper non-informative 
priors, partly out of convenience and partly 
in order to be more objective, letting the data 
show the way. The purists, on the other hand, 
insist on proper priors.

From a computational point of view, the propriety 
of the prior has important implications. Obviously, 
one cannot compute prior predictive p-values, for instance, 
if one cannot sample form the prior. This also applies 
to our calibrated posterior predictive p-values.
We would however argue that the important distinction 
should not be between mathematically proper or improper 
prior distributions, but rather between reasonable 
and unreasonable ones. 
By a `reasonable prior' we mean a distribution under which 
it makes sense to sample. As an example of a mathematically 
proper, but highly unreasonable, prior we might turn
to the hazard rate analysis example in 
Gelman, Meng and Stern (1996). 
To carry out Bayes analysis for the survival data 
in question one needs to model the probabilities 
$\theta_i$ of a person not surviving a year $35+i$, 
given that the person survived year $34+i$, 
for $i=1,\ldots,30$. Their prior is a uniform on $[0,1]^{30}$, 
constrained by the assumption that the sequence 
$\theta_1,\ldots,\theta_{30}$ is increasing and convex.
Without looking at the specific data of this application,
we know that $\theta_{30}$ is well below .5, 
most likely below .1. But under the chosen prior, 
$\Pr\{\theta_{30}>0.5\} < 10^{-9}$; even more 
diabolically, under the Gelman, Meng, Stern prior 
nearly every human being would die before the age of 65,
and very few would celebrate their 60th birthday. 

It therefore makes no sense to sample under this prior. 
Our calibration method is accordingly not applicable
for this application, and the problem of defining the 
`true surprise' of a computed ppp value has no obvious solution.
This is a case where we might want to apply our 
`what if' analysis of Section 4.3. 
We may keep the unreasonable prior of Gelman, Meng and Stern 
as our fitting prior, which means using it for computing 
the posterior probability distribution,
and also for the $\ppp$ value. When calibrating this 
$\ppp$ value, however, we would use a much more
informative prior, e.g.~by restricting the uniform prior
on increasing convex vectors in $[0,1]^{30}$ 
by $\theta_{30}\in[0.03,0.07]$, if we consider this 
a likely range. One may ask why, then, do we not use 
this improved prior throughout the analysis? Firstly, 
we may not wish to include this subjective piece of information 
in our otherwise objective analysis of the data.
Applying a subjective prior in the process of testing 
the sanity of the model is quite different from 
using it as a basis for posterior inference. 
Secondly, there may be a risk that our sampling prior is too narrow,
failing to capture the `true' parameter value.
We may therefore prefer to fit the model under a prior 
that is vague, so that the data get the chance of 
showing the way. Similarly, we may want the sampling prior 
used for calibrating the $\ppp$ value to be on the narrow 
side, in order to ensure that we do not calibrate our 
$\ppp$ value against nonsense.

\section
\centerline{\bf 10. Related themes and concluding remarks}

\hop 
We end our article with a list of comments
and indication of themes, some of which might 
warrant further research efforts. 

\subsection
{\sl 10.1. Calibrating the prior through the calibrated ppp.}
Eliciting and fine-tuning priors remains of course 
a difficult task, even for experienced Bayesian 
statisticians. There are often situations where 
the statistician may translate prior knowledge 
into a reasonably secure centre point, say a 
`prior guess' equal to $\theta_0$, but where setting 
the appropriate precision level is far from 
clear-cut. In such situations a scheme or way 
of thinking not infrequently followed is to 
try out different precision levels, from 
`reasonably precise' to `quite non-informative',
and make do with a level of spread that balances
the two desiderata of not trusting the prior guess
too much and at the same time not tolerating 
a serious conflict or clash between prior and data.
Some `prior sample size' elicitations are indirectly
of such a form. 

The procedure just described is somewhat ad hoc, 
of course, and may not be easy to follow or formalise 
in practice. The cppp mechanism offers a venue leading 
to a precise version of this idea, however. The proposal 
is to monitor the $\cppp(y^\obs)$ as a function 
of the spread or precision parameter involved, 
and in the end use the most conservative prior 
that still does not clash with data, in the sense 
of having for example $\cppp(y^\obs)=.10$, 
but not lower. 
As a mundane illustration of this cppp-induced 
conservative prior, study again the normal--normal 
setup of Section 2 and Section 4.1. Assume the 
statistician has selected a secure prior guess 
parameter $\theta_0$ for $\theta$, but is not 
yet certain about the choice of the prior's
spread parameter $\sigma_0$. From (4.4), 
the idea above amounts to selection $\sigma_0$ to have 
$${n(\bar y^\obs-\theta_0)^2/\sigma^2\over 1+n\sigma_0^2/\sigma^2}
  =1.645^2, \quad {\rm or} \quad
  \sigma_0={\sigma\over \rootn}
  \Bigl({z_n^\obs\over 1.645^2}-1\Bigr)^{1/2}, $$
where $z_n^\obs=n(\bar y^\obs-\theta_0)^2/\sigma^2$ 
is the usual test statistic for testing the null hypothesis 
that $\theta=\theta_0$. This formula would be used when 
$z_n^\obs\ge 1.645^2$, i.e.~when the test rejects $\theta_0$
at significance level $0.10$. In cases where 
the $\theta=\theta_0$ hypothesis is accepted by
the data, the scheme above would allow a sharp prior 
at $\theta_0$, i.e.~setting $\sigma_0$ to a small value. 


We stress that the idea is quite general and might 
be used in situations much less clear-cut than tha above,
e.g.~in semi- and nonparametric contexts. One 
type of application would be to nonparametric 
setups involving Dirichlet or Beta processes, 
where the statistician knows where to centre these
but is unsure about the concentration parameters. 

\subsection
{\sl 10.2. Detection power.}
There is not room here for properly discussing 
the detection power of the cppp assessors. 
This clearly depends on aspects of the situation
at hand, including the discrepancy function 
$D(y,\theta)$. For the normal--normal setup of 
Section 2 it is not difficult to study the distribution
of $U=\ppp(Y)$ under various conditions different 
from those implied by the prior and the model;
we will report elsewhere on some such findings. 
While the power may be satisfactory for some 
sets of alternative combinations of prior and model 
there will remain types of alternatives that are difficult 
to detect, with any given $D(y,\theta)$.  
For detecting special types of violations one might 
therefore need to devise corresponding special 
discrepancy functions. 

\subsection
{\sl 10.3. cppp as a p-value.}
One way of viewing our cppp construction is that 
the $\ppp(y^\obs)$ of (1.1), albeit clearly having 
Bayesian interpretation and inspiration, is nothing
but a test statistic. Since statisticians compute it,
they wish directly or indirectly to assess its significance, 
which amounts to comparing it to its null distribution. 
This is in effect what the cppp operation does. 

\subsection
{\sl 10.4. Tail versus height.}
When attempting to assess the degree of surprise 
in an observation $U=u$, one might compute tail areas
for the relevant null distribution, 
or a suitable ratio of densities  $h(u)/g(u)$.
We have focussed on the first general direction, 
partly because is would be difficult to find good general 
candidates for the $h(u)$ in question, and partly since the null
densities $g(u)$ have been seen to be so extreme,
cf.~Figure 4.1.  

\subsection
{\sl 10.5. General hierarchical models.}
In complicated hierarchical models it is easy to lose
track of the different implications of a many-levelled
prior. Ordinary assessment methods may not work well
in such cases, see e.g.~comments made in 
Lu, Hodges and Carlin (2004).
The cppp methods of 
our article may be generalised to various hierarchical
models, and we believe they may be useful for
screening out unfortunate combinations of 
prior and model there, when employed with appropriate
discrepancy functions. 

\medskip
{\bf Acknowledgements.}
The article has benefitted from constructive 
suggestions from the Associate Editor, 
Editor Frank Samaniego, Alan Gelfand, Bent Natvig,
Geir Storvik, and two anonymous reviewers. 
The research of Dahl and Steinbakk has been
partly supported by the Norwegian Science Council
under the BeMatA programme `Evaluation of Hierarchical Models',
funded by the Norwegian Research Council, led by Hjort. 


\def\JRSS{Journal of the Royal Statistical Society}
\def\JASA{Journal of the American Statistical Association}
\def\SS{Statistical Science} 
\def\JSPI{Journal of Statistical Planning and Inference}
\def\AoS{Annals of Statistics} 

\section
\centerline{\bf References}

\def\ref#1{{\noindent\hangafter=1\hangindent=20pt
  #1\smallskip}}          
\parindent0pt
\baselineskip11pt
\parskip3pt 

\medskip

\ref{%
Bayarri, M.J.~and Berger, J.O. (2000).
P values in composite null models 
[with discussion and a rejoinder]. 
{\sl \JASA} {\bf 95}, 1127--1142.}

\ref{%
Bayarri, M.J.~and Berger, J.O. (2004).
The interplay of Bayesian and frequentist analysis.
{\sl \SS} {\bf 19}, 58--80.}

\ref{%
Bayarri, M.J.~and Castellanos, M.E. (2004). 
Bayesian checking of hierarchical models. 
Technical report, University of Valencia.}


\ref{%
Billingsley, P. (1968).
{\sl Convergence of Probability Measures.}
Wiley, Singapore.}

\ref{%
Box, G.E.P. (1980).
Sampling and Bayes' inference in scientific modelling and robustness.
{\sl \JRSS} {\bf A 143}, 383--430.}

\ref{%
Brooks, S.P., Catchpole, E.A.~and Morgan, B.J.T. (2000).
Bayesian animal survival estimation.
{\sl \SS} {\bf 15}, 357--376.}

\ref{%
Claeskens, G.~and Hjort, N.L. (2003).
The focused information criterion
[with discussion contributions and a rejoinder].
{\sl \JASA} {\bf 98}, 900--916 and 938--945.}

\ref{%
Clyde, M.~and George, E. (2004).
Model uncertainty. 
{\sl \SS} {\bf 19}, 81--94.}



\ref{%
Dey, D.K., Gelfand, A.E., Swartz, T.B.~and Vlachos, P.K. (1998).
A simulation-intensive approach for checking hierarchical models.
{\sl Test} {\bf 7}, 325--346.}


\ref{%
Gelfand, A.E.~and Dey, D.K. (1994).
Bayesian model choice: Asymptotics and exact calculations.
{\sl \JRSS} {\bf B 56}, 501--514.}

\ref{%
Gelfand, A.E.~and Ghosh, S.K. (1998).
Model choice: A minimum posterior predictive approach.
{\sl Biometrika} {\bf 85}, 1--11.}

\ref{%
Gelfand, A.E.~and Wang, F. (2002).
A simulation based approach to sample size determination
under a given model and for separating models.
{\sl \SS} {\bf 17}, 193--208.}

\ref{%
Gelman, A., Meng, X.-L.~and Stern, H. (1996).
Posterior predictive assessment of model fitness
via realized discrepancies [with discussion]. 
{\sl Statistica Sinica} {\bf 6}, 733--807.}

\ref{%
Gelman, A., Carlin, J.B., Stern, H.S.~and Rubin, D.B. (2004).
{\sl Bayesian Data Analysis} (2nd ed.).
Chapman and Hall/CRC Press, Boca Raton, Florida.}

\ref{%
Good, I.J. (1983).
{\sl Good Thinking: The Foundations of Probability
and Its Applications.}
University of Minnesota Press, Minneapolis.}  

\ref{%
Gutti\'errez-Pe{\~n}a, E.~and Walker, S.G. (2001).
A Bayesian predictive approach to model selection.
{\sl \JSPI} {\bf 93}, 259--276.}

\ref{%
Guttman, I. (1967).
The use of the concept of a future observation
in goodness-of-fit problems.
{\sl \JRSS} {\bf B 29}, 83--100.}

\ref{%
O'Hagan, T. (2003). 
HSSS model criticism [with discussion].
In {\sl Highly Structured Stochastic Systems}
(eds.~P.J.~Green, N.L.~Hjort and S.~Richardson),
Oxford University Press, Oxford, pp.~423--453.}




\ref{%
Hjort, N.L. (2003a).
Catriona LeMay Doan: Going for Gold [book review]. 
{\sl SpeedSkating World} {\bf 8}, 14--15.}

\ref{%
Hjort, N.L. (2003b).
Topics in nonparametric Bayesian statistics [with discussion].
In {\sl Highly Structured Stochastic Systems}
(eds.~P.J.~Green, N.L.~Hjort and S.~Richardson),
Oxford University Press, Oxford, pp.~455--478.}

\ref{%
Hjort, N.L. (2005).
The Bolshoi of speedskating 
[report on the World Championships]. 
{\sl SpeedSkating World} {\bf 10}, 14--17.}

\ref{%
Hjort, N.L.~and Claeskens, G. (2003).
Frequentist model average estimators
[with discussion contributions and a rejoinder].
{\sl \JASA} {\bf 98}, 879--899 and 938--945.}

\ref{%
Hjort, N.L.~and Ongaro, A. (2005). 
Exact inference for random Dirichlet means. 
{\sl Statistical Inference for Stochastic Processes}
(to appear).}

\ref{%
Kadane, J.B.~and Lazar, N.A. (2004).
Methods and criteria for model selection.
{\sl \JASA} {\bf 99}, 279--290.}

\ref{%
Kass, R.E.~and Raftery, A.E. (1995).
Bayes factors.
{\sl \JASA} {\bf 90}, 773--795.}

\ref{%
Lebreton, J.-D., Burnham, K.P., Clobert, J.~and Anderson, D.R. (1992).
Modeling survival and testing biological hypotheses using marked animals:
a unified approach with case studies. 
{\sl Ecological Monographs} {\bf 62}, 67--118.} 

\ref{%
Lehmann, E.L. (1983).
{\sl Theory of Point Estimation.}
Wiley, New York.}

\ref{%
van der Linde, A. (2004).
On the association between a random variable and an observable.
{\sl Test} {\bf 13}, 85--111.}

\ref{%
Lu, H., Hodges, J.S.~and Carlin, B.P. (2004).
Measuring the complexity of generalized linear hierarchical models.
Technical report.}

\ref{%
Marden, J.I. (2000).
Hypothesis testing: from p values to Bayes factors.
{\sl \JASA} {\bf 95}, 1316--1320.}

\ref{%
Meng, X.-L. (1994).
Posterior predictive p-values.
{\sl \AoS} {\bf 22}, 1142--1160.}

\ref{%
Robert, C.P. (2001).
{\sl The Bayesian Choice} [2nd ed.].
Springer, New York.}

\ref{%
Robins, J.M., van der Vaart, A.~and Ventura, V. (2000).
Asymptotic distribution of p values in composite null models
[with discussion and a rejoinder]. 
{\sl \JASA} {\bf 95}, 1143--1156.}

\ref{%
Rubin, D.B. (1984). 
Bayesianly justifiable and relevant frequency calculations
for the applied statistician. 
{\sl \AoS} {\bf 12}, 1151--1172.} 

\ref{%
Sinharay, S.~and Stern, H.S. (2003).
Posterior predictive model checking in hierarchical models.
{\sl \JSPI} {\bf 111}, 209--221.}

\ref{%
Smith, A.F.M.~and Spiegelhalter, D.J. (1980).
Bayes factors and choice criteria for linear models.
{\sl \JRSS} {\bf B 42}, 213--220.}

\ref{%
Spiegelhalter, D.J., Best, N.G., Carlin, B.P.~and van der Linde, A. (2002).
Bayesian measures of model complexity and fit 
[with discussion and a rejoinder].
{\sl \JRSS} {\bf B 64}, 583--639.}

\bye